\documentclass{article}

\usepackage[left=2.5cm,right=2.5cm]{geometry}
\usepackage{graphicx}%
\usepackage{multirow}%
\usepackage{amsmath,amssymb,amsfonts}%
\usepackage{amsthm}%
\usepackage{mathrsfs}%
\usepackage[title]{appendix}%
\usepackage{xcolor}%
\usepackage{textcomp}%
\usepackage{manyfoot}%
\usepackage{booktabs}%
\usepackage{algorithm}%
\usepackage{algorithmicx}%
\usepackage{algpseudocode}%
\usepackage{listings}%
\usepackage{subcaption} %
\usepackage{siunitx} %
\usepackage{breakurl}
\usepackage{multirow}
\usepackage{authblk}
\usepackage{hyperref}

\newcommand{\friction}{\alpha}
\newcommand{\velocity}{\mathbf{v}}
\newcommand{\temperature}{T}
\newcommand{\height}{h}

\newcommand{\heatflux}{q_{\rm{geo}}}
\newcommand{\heatfluxG}{q_{geo}^{\rm{G}}}
\newcommand{\heatfluxSR}{q_{geo}^{\rm{SR}}}

\newcommand{\strainratex}{\dot{\varepsilon}_{\rm{HO}, 1}}
\newcommand{\strainratey}{\dot{\varepsilon}_{\rm{HO}, 2}}
\newcommand{\strainrate}{\dot{\varepsilon}}

\newcommand{\ddx}[1]{\frac{\partial #1}{\partial x}}
\newcommand{\ddy}[1]{\frac{\partial #1}{\partial y}}
\newcommand{\ddz}[1]{\frac{\partial #1}{\partial z}}

\newcommand{\tint}{t_{0}}
\newcommand{\tfin}{t_{f}}

\newcommand{\correlation}[1]{\rho_{#1}}
\newcommand{\levelvariance}[2]{\overline{\sigma}_{#1}^2}

\makeatletter
\def\botrule{\noalign{\ifnum0=`}\fi
  \hrule \@height 3pt \@width 0pt
  \hrule \@height 0.75\p@ 
  \hrule \@height 3pt \@width 0pt
  \futurelet\@tempa\@xhline}
\makeatother

\begin{document}

\title{Multifidelity Uncertainty Quantification for Ice Sheet Simulations}

\author[1]{Nicole Aretz}
\author[1]{Max Gunzburger}
\author[2]{Mathieu Morloghem}
\author[1]{Karen Willcox}

\affil[1]{{Oden Institute for Computational Engineering and Sciences}, {University of Texas at Austin}, {{201 E 24th St}, {Austin}, {78712}, {Texas}, {USA}}}

\affil[2]{{Department of Earth Sciences}, {Dartmouth College}, {{Hanover}, {03755}, {New Hampshire}, {USA}}}

\maketitle


\abstract{
Ice sheet simulations suffer from vast parametric uncertainties, such as the basal sliding boundary condition or geothermal heat flux. 
Quantifying the resulting uncertainties in predictions is of utmost importance to support judicious decision-making, but high-fidelity simulations are too expensive to embed within uncertainty quantification (UQ) computations. 
UQ methods typically employ Monte Carlo simulation to estimate statistics of interest,  which requires hundreds (or more) of ice sheet simulations.  
Cheaper low-fidelity models are readily available (e.g., approximated physics, coarser meshes), but replacing the high-fidelity model with a lower fidelity surrogate introduces bias, which means that UQ results generated with a low-fidelity model cannot be rigorously trusted. 
Multifidelity UQ retains the high-fidelity model but expands the estimator to shift computations to low-fidelity models, while still guaranteeing an unbiased estimate. 
Through this exploitation of multiple models, multifidelity estimators guarantee a target accuracy at reduced computational cost. 
This paper presents a comprehensive multifidelity UQ framework for ice sheet simulations. 
We present three multifidelity UQ approaches---Multifidelity Monte Carlo, Multilevel Monte Carlo, and the Best Linear Unbiased Estimator---that enable tractable UQ  for continental-scale ice sheet simulations.
We demonstrate the techniques on a model of the Greenland ice sheet to estimate the 2015-2050 ice mass loss, verify their estimates through comparison with Monte Carlo simulations, and give a comparative performance analysis. 
For a target accuracy equivalent to \SI{1}{mm} sea level rise contribution at \SI{95}{\%} confidence, the multifidelity estimators achieve computational speedups of two orders of magnitude.
}


\section{Introduction}\label{sec:introduction}

\begin{figure*}
    \centering
    \includegraphics[width=\textwidth]{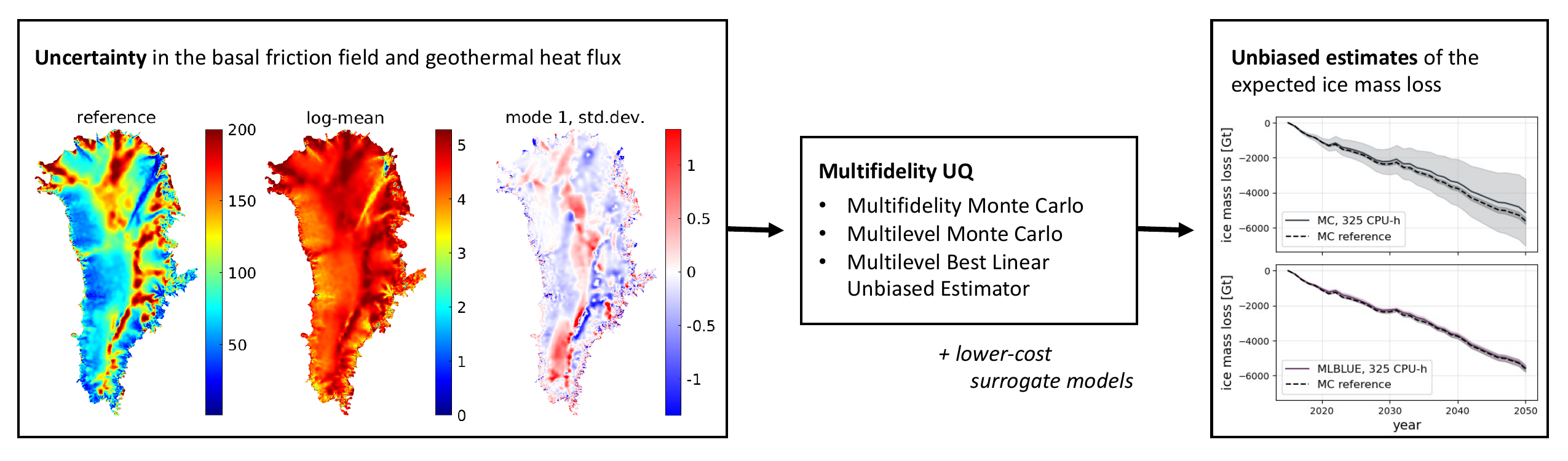}
    \caption{Multifidelity UQ framework}
    \label{fig:intro}
\end{figure*}

Sea level rise is impacting both coastal ecosystems and our societies. Mass loss from ice sheets is becoming a major driver of sea level change, and large-scale simulations of the Greenland and Antarctic ice sheets have a central role when evaluating policies to combat climate change. 
Since these simulations must be over time horizons spanning decades and centuries, modeling decisions can significantly influence the predicted ice mass change and thus the predicted contribution to sea level rise.
Even among state-of-the-art high-fidelity ice sheet models, as illustrated by model intercomparison studies \cite{ismip6,goelzer2018design, seroussi2020ismip6, seroussi2019initmip}, expert opinions on modeling parameterizations and datasets differ, and there is high variance in ice mass loss projections.
To provide effective support to decision-making, it is paramount to quantify the uncertainty associated with these simulation-based projections, yet doing so is computationally prohibitive because it would require an ensemble of high-fidelity simulations to be run over many different modeling choices. In this paper we lay out a multifidelity uncertainty quantification (UQ) framework that addresses this challenge.

Model intercomparison studies have a long tradition in glaciology, including in model validation \cite{huybrechts1996eismint,pattyn2008benchmark,payne2000results}, verification based on historical data \cite{goelzer2018design, seroussi2019initmip}, and for high-fidelity projections \cite{ismip6, seroussi2020ismip6, hock2019glaciermip}.
Moreover, parametric uncertainties in ice sheets are studied extensively and explored often through sensitivity analyses, e.g., 
\cite{van2012sensitivity, rogozhina2012effects, larour2012ice} for the geothermal heat flux, or \cite{barnes2022predictive, zhao2018basal,cheng2021sensitivity} for the basal friction.
Acknowledging parametric uncertainty and its importance for high-fidelity projections, most ice sheet models infer the basal friction field from observational data of the surface ice velocity in a deterministic \cite{morlighem2010spatial, morlighem2013inversion} or more recently Bayesian \cite{isaac2015scalable, babaniyi2021inferring, brinkerhoff2021constraining} inverse problem; other parameters like the geothermal heat flux \cite{shapiro2004inferring, greve2019geothermal} or the basal topography \cite{morlighem2017bedmachine,brinkerhoff2016bayesian, guan2018inferring} are inferred once and then distributed throughout the community through datasets.
While these are examples of \textit{inverse} UQ, where data is used to reduce the uncertainty in a parameter, in this paper we are primarily concerned with \textit{forward} UQ, where the goal is to quantify the influence of parametric uncertainty on an output of interest (OoI) such as the predicted ice mass loss (see Fig.~\ref{fig:intro}).

Forward UQ is challenging for computationally expensive ice sheet simulations since typically a large ensemble of projections is required to approximate the statistics (e.g., the mean) of the OoI under a given probabilistic description of parametric uncertainties.
In \cite{bamber2013expert, oppenheimer2016expert}, expert judgements were pooled and formalized to quantify the uncertainty in existing ice sheet projections.
In \cite{machguth2008exploring}, a mass-balance model of so-called ``intermediate complexity" was used in a Monte Carlo (MC) estimation with 5,000 samples, but the projection was limited to a time horizon of 400 days and a glacier of approximately \SI{16}{km^2}.
Other studies have balanced required ensemble sizes and computational costs by accepting a \SI{10}{km} resolution \cite{golledge2015multi, ritz2015potential}, using hybrid or approximated physics models \cite{deconto2016contribution, edwards2019revisiting}, or built probabilistic emulators \cite{little2013upper, bulthuis2019uncertainty, pollard2016large, aschwanden2019contribution, edwards2021projected} trained on high-fidelity simulations.
A drawback in using these modeling simplifications is the model bias  introduced by replacing the high-fidelity ice sheet model with a cheaper surrogate model. 
In doing so, there is no guarantee that the UQ results will reflect what would have been obtained using the most trusted high-fidelity model.
Multi-model ensemble studies have also been employed to compare the deterministic projections of several high-fidelity models, each built to reflect their modeller's best domain expertise.
However, because these high-fidelity models are so expensive, only small ensembles are possible.
Moreover, most ensemble studies do not account for the uncertainties in each individual model.
Multifidelity UQ methods can overcome these challenges:
Formal UQ at a desired target accuracy is achieved by leveraging surrogate models to reduce the computational cost, but in a structured way that guarantees a statistically unbiased estimate of the high-fidelity OoI statistic.

Multifidelity UQ methods exploit the correlation between the high-fidelity model and less accurate but computationally cheaper surrogate models to construct an unbiased estimator of the high-fidelity expectation.
Compared to MC sampling, multifidelity estimators achieve an improved accuracy for any prescribed computational budget (or equivalently, they achieve a given target accuracy with a reduced computational budget).
In this paper we present three predominant multifidelity UQ methods, and highlight their applicability for ice sheet simulations:
the Multifidelity Monte Carlo (MFMC) method \cite{ng2014multifidelity,Peherstorfer2016b}, the Multilevel Monte Carlo (MLMC) method \cite{heinrich2001multilevel,giles2008multilevel}, and the Multilevel Best Linear Unbiased Estimator (MLBLUE) \cite{Schaden2020}.
Our paper has three objectives:
    (1)~To establish the necessity of unbiased UQ for ice mass loss projections 
    by showing the errors that can be incurred by using approximate models without a formal multifidelity framework;
    (2)~to demonstrate the efficiency of multifidelity UQ methods for computationally expensive glaciology simulations;
    and (3)~to provide a simple but flexible algorithmic framework to implement, use, and interpret MFMC, MLMC, and MLBLUE on any ice sheet code without prior experience in UQ or surrogate modeling.
To emphasize the relevance and applicability of our work, we employ a community ice sheet code --- the Ice-sheet and Sea-level System Model (ISSM, \cite{issm}) --- and follow the projection protocols \cite{nowicki2020experimental} of the Ice Sheet Model Intercomparison Project (ISMIP6, \cite{nowicki2016ice}) contribution to Coupled Model Intercomparison Project Phase 6 (CMIP6, \cite{eyring2016overview}), with the goal to make the transfer of techniques to similar applications as easy as possible.

In Section~\ref{sec:setting}, we describe our high-fidelity model of the Greenland ice sheet, describe uncertainties in the basal friction and geothermal heat flux fields introduced by different data sets, and show the large variance these cause in the high-fidelity projections.
We also discuss different types of surrogate models, and illustrate their model bias.
In Section \ref{sec:methods} we provide primers to the MFMC, MLMC, and MLBLUE methods, each with a detailed algorithmic description to facilitate their implementation.
In Section \ref{sec:results}, we demonstrate these methods for estimating the expected ice mass loss for the year 2050, and discuss the benefits of multifidelity UQ for ice sheet simulations.

\section{Model of the Greenland Ice Sheet}\label{sec:setting}

This section sets up the high-fidelity model of the Greenland ice sheet, which we use to demonstrate the necessity and challenges of UQ for ice sheet simulations.
Our high-fidelity model comprises governing equations for ice temperature, velocity and thickness (Section~\ref{sec:high-fidelity}).
We describe uncertainties in the basal friction field and the geothermal heat flux that stem from the variety of available datasets in the literature (Section~\ref{sec:uncertainties}).
The predictive uncertainty thus reflects the influence that datasets have on model simulations.
This section also provides brief descriptions of some of the many surrogate models readily available in ice sheet codes (Section~\ref{sec:surrogates}), including coarse-grid approximations, simplified-physics models, and statistical emulators.

\subsection{High-Fidelity Model}\label{sec:high-fidelity}

We consider a model of the Greenland ice sheet that consists of three coupled nonlinear partial differential equations (PDEs), namely 
\begin{itemize}
\item a thermal model, which governs changes in the temperature $\temperature$ within the ice sheet, based on the conservation of energy;
\item a dynamical model with nonlinear rheology for the velocity vector $\velocity = (v_x, v_y, v_z)^{\top}$, which governs the motion of the ice sheet, based on the conservation of momentum;
\item a mass transport model, which governs changes in the thickness $\height$ of the ice sheet, based on the conservation of mass.
\end{itemize}
Out of these variables, the temperature $\temperature$ and the velocity components $v_x$, $v_y$, and $v_z$ in $x$-, $y$-, and $z$-direction are each three-dimensional fields, defined on the ice sheet domain $\Omega(t)$;
the ice thickness $\height$ is two-dimensional, defined on the horizontal extent $\Omega_{\rm{2D}}$ of the ice sheet, and determines how the ice geometry $\Omega(t)$ evolves in time $t$ vertically.

The specific three-part system we consider is a simplification of more sophisticated ice sheet models\footnote{Removing the simplifications would result in even better performance of the multifidelity UQ estimation approaches discussed in this paper, but then the reference high-fidelity UQ results would become computationally prohibitive, preventing us from analyzing the performance of the multifidelity approaches.} that incorporate additional ice sheet features such as ice--ocean interactions and basal hydrology.
Despite these simplifications, the system we consider is already at a scale where UQ tasks are computationally demanding, and suffices to demonstrate how multifidelity UQ techniques can obtain trusted estimates of statistical OoIs at a much lower computational cost than MC estimation.

\subsubsection{Geometry and Initialization}

We model the main part of the Greenland ice sheet under the shared economic pathway scenario SSP1-2.6 (\cite{RN1,RN3}, chapter 1.6), which is a low emission scenario.
For our atmospheric forcing, we follow the experimental protocol \cite{nowicki2020experimental} of the ISMIP6 Greenland study \cite{ismip6}.
Consequently, the projections for our multifidelity UQ study start in the year 2015, and we perform a spin-up run from our model initialization in $\tint:=1989$ to obtain the 2015 initial conditions (described below).
For facilitating comparisons with Monte Carlo sampling, it suffices that we limit our projections to the years $2015$ to $\tfin:=2050$, though the methodology applies analogously to longer projection regimes.

For our domain outline, we traced any ice of thickness greater than \SI{5}{m} reported in \cite{BedMachineV5, morlighem2017bedmachine}, removed all disconnected parts (e.g., islands, numerical artifacts) and major bottlenecks, and smoothed the obtained outline.
Our final outline encompasses approximately 86.64 \% of the ice-covered area in \cite{howat2014greenland} (version 2, with coastline from Jeremie Mouginot) and approximately 99.59 \% of the total ice mass in \cite{BedMachineV5, morlighem2017bedmachine}.
Using the 1995--2015 averaged surface ice velocity from \cite{MEaSUREs-Vel_v1} and the ice thickness \cite{BedMachineV5, morlighem2017bedmachine} for reference, we create our high-fidelity mesh with a resolution varying from \SI{100}{m} to \SI{15}{km}.
We denote this domain by $\Omega_{\rm{2D}}$, and keep it fixed throughout all simulations,

To obtain the basal topography $\Gamma_{\rm{b}}$, we interpolate the ice base from \cite{BedMachineV5, morlighem2017bedmachine} onto our mesh.
Similarly, for the surface topography $\Gamma_{\rm{s}}(\tint)$, we interpolate the ice altitude data from \cite{howat2014greenland} (version 2).
Following \cite{morlighem2017bedmachine}, these initializations give us a geometry consistent with the year 2009.
For any $\tint \le t \le 2009$, we thus keep the geometry fixed, with a constant ice thickness $\height(t) = \Gamma_{\rm{s}}(\tint) - \Gamma_{\rm{b}}$.
Starting from the year $t=2009$, $\height(t)$ obeys the ice thickness equations defined in Section \ref{sec:thickness}, and the ice surface topography $\Gamma_{\rm{s}}(t) = \Gamma_{\rm{b}} + \height(t)$ evolves accordingly.
For 3D variables, the 2D domain $\Omega_{\rm{2D}}$ and its mesh are extruded with five layers, starting from the ice base $\Gamma_{\rm{b}}$ upwards to the ice surface $\Gamma_{\rm{s}}(t)$.
We denote the obtained 3D domain by $\Omega(t)$, using the variable $t$ to stress its variability in time.

To initialize our model in $\tint=1989$, we compute the steady-state equilibrium of the ice temperature $\temperature$ and velocity $\velocity$ with CNRM-CM6-1 1960--1989 reference forcing \cite{voldoire2018cnrm}.
This initialization mimics that the Greenland ice sheet is believed to have been in steady state in the 1960--1989 time period \cite{mouginot2019forty, rignot2006changes}.
We then run our model using the equations described in the following subsections,
until the start year $t=2015$ prescribed in \cite{nowicki2020experimental} to obtain the initial conditions of temperature $\temperature$, velocity $\velocity$, and ice thickness $\height$ used in our predictive runs.

\subsubsection{Stress balance model}\label{sec:velocity}

There exists a hierarchy of ice sheet dynamical models that can be ordered according to decreasing physical fidelity and which, for the most part, also possess decreasing computational simulation costs. At the top of the hierarchy is the most generally accepted (with respect to physical fidelity) {\em full-Stokes model} (FS).
However, for a given ice geometry $\Omega(t)$ and a given temperature field $\temperature(t)$, the numerical solution of the FS model is challenging and expensive for several reasons, which include the usage of fine grids and long time horizons; stable and sufficiently accurate spatial and temporal discretization choices (e.g., finite element and time-stepping schemes, respectively); and the need to solve for four field variables over the 3D domain (three velocity components and the pressure).
As a result, few ice sheet codes support the FS equations.
In significantly greater use by practitioners is the {\em higher-order model}\footnote{Here, the nomenclature ``higher order'' is relative to even more simplified ice sheet models such as the shallow-shelf and shallow-ice models; see e.g., Section \ref{sec:surrogates}, and \cite{greve2009dynamics}.} (HO),
which is an ice sheet dynamical model representing the next lower level of physical fidelity from FS. 
For reference, in the ISMIP6 Greenland study \cite{ismip6}, none of the 21 participating models solved the FS system, but seven solved the HO equations. 
The remaining 14 groups opted for further physical approximations in the form of the Shallow-Shelf or Shallow-Ice Approximations, or their hybrid combination (\cite{ismip6}, Table A1).

The HO model is derived as a simplification of the FS model by taking advantage of the small vertical-to-horizontal aspect ratio of the ice sheet. 
The net result is that the horizontal gradients of vertical velocities are neglected compared to vertical gradients of horizontal velocities, and bridging effects are also neglected. 
The resulting HO model is then given by the system of PDEs
\begin{align}
    \nabla \cdot (2\mu\strainratex) = \rho_{\rm{ice}} g \frac{\partial s(t)}{\partial x} \label{hoxx}\\
    \nabla \cdot (2\mu\strainratey) = \rho_{\rm{ice}} g \frac{\partial s(t)}{\partial y} \label{hoyy}
\end{align}
for $\mathbf{x}=(x, y, z)\in\Omega(t)$ and $\tint< t \le \tfin$, and where $\rho_{\rm{ice}} = \SI{917}{kg/m^{3}}$ is the ice density, $g = \SI{9.81}{m/s^{2}}$ is the gravitational acceleration, $s(x,y; t)$ is the altitude of the surface $\Gamma_{\rm{s}}(t)$ vertically above the point $\mathbf{x}=(x,y,z)$, and the strain rates $\strainratex$, $\strainratey$ take the form
\begin{align*}
    \strainratex = \left(
        \begin{array}{c}
          \displaystyle  2 \ddx{v_x} + \ddy{v_y} \\
           \displaystyle \frac12 \ddy{v_x} + \frac12 \ddx{v_y} \\
           \displaystyle \frac12 \ddz{v_x}
        \end{array}
    \right), \qquad
    \strainratey = \left(
        \begin{array}{c}
          \displaystyle  \frac12 \ddy{v_x} + \frac12 \ddx{v_y} \\
          \displaystyle  \ddx{v_x} + 2\ddy{v_y} \\
          \displaystyle  \frac12 \ddz{v_y}
        \end{array}
    \right).
\end{align*}
\noindent The effective ice viscosity $\mu$ follows the generalized Glen's flow law
\begin{align}\label{eq:Glen}
    \mu =  \frac{B(\temperature)}{\sqrt{2}} \left\| (\strainratex, \strainratey) \right\|^{-\frac{n-1}{n}}
\end{align}
where the Glen's flow law exponent $n$ is typically taken as $n = 3$. 
To determine the local ice rigidity $B(\temperature)$ for the ice temperature $\temperature(t)$, we apply the temperature relationship provided by \cite[p. 97][]{Paterson1994}.

The system
\eqref{hoxx} and \eqref{hoyy} 
is subject to homogeneous zero-Neumann boundary conditions on the surfaces $\Gamma_{\rm{air}}(t)$ exposed to air, i.e.,
\begin{align*}
    \strainratex \cdot \mathbf{n} &= 0, & \strainratey \cdot \mathbf{n} &= 0
\end{align*}
for $\mathbf{x}=(x,y,z) \in \Gamma_{\rm{air}}(t)$, and
where $\mathbf{n} = (n_x, n_y, n_z)$ is the outward pointing unit normal of $\Omega(t)$.
The stress balance with the atmospheric pressure has been neglected because that pressure is negligible in the setting we consider.
In contrast, at the ice-water interface $\Gamma_{\rm{w}}$ the water pressure gets applied through
\begin{equation*}
    \begin{aligned}
        &2 \mu \strainratex \cdot \mathbf{n} = f_{\rm{w}} \mathbf{n}_x, \quad 2 \mu \strainratey \cdot \mathbf{n} = f_{\rm{w}} \mathbf{n}_y
    \end{aligned}
\end{equation*}
for $\mathbf{x}=(x,y,z) \in \Gamma_{\rm{w}}$, and with 
\begin{align*}
    f_{\rm{w}}(\mathbf{x}) = \rho_{\rm{ice}} g (s(x,y; t) - z) + \rho_{\rm{w}} g \min\{z, 0\}
\end{align*}
where $\rho_{\rm{w}} = \SI{1023}{km/m^3}$ is the ocean water density.
At the basal boundary $\Gamma_{\rm{b}}$, we prescribe sliding boundary conditions
\begin{equation*}
    \begin{aligned}
        2 \mu \strainratex \cdot \mathbf{n} &= -\friction^2 N\;v_x, 
        &2 \mu \strainratey \cdot \mathbf{n} &= -\friction^2 N\;v_y 
    \end{aligned}
\end{equation*}
for $\mathbf{x}=(x,y,z) \in \Gamma_{\rm{b}}$, and
where $\friction$ is the basal friction field, and $N=\rho_{\rm{ice}} g h + \rho_{\rm{water}} g b$, with $b$ denoting the altitude of the basal topography $\Gamma_{\rm{b}}$, is the effective pressure, following Budd's friction law \cite{Budd1979}.
We make the simplifying assumption that all ice is grounded such that $\friction$ is constant in time.
Note that floating ice is not essential to our considered scenario as this assumption only affects \SI{0.22}{\%} of the domain.

Since the basal friction field $\friction$ is only indirectly observable, we model it as a random variable, see Section \ref{sec:uncertainties}.
However, in our reference parameterization used for the spin-up and control runs, we choose $\friction_{\rm{ref}}$ by  minimizing the misfit between observed and simulated surface velocity data, described by the cost function
\begin{equation}\label{eq:hf:costfunction}
    \begin{aligned}
    &J(\friction; \velocity_{\rm{obs}}) 
    = 180 \int_{\Gamma_{\rm{s}}} \|\velocity(\friction) - 
    \velocity_{\rm{obs}}\|^2  dS  
    + 0.6 \int_{\Gamma_{\rm{s}}} \left(\log \left( \frac{\|\velocity(\friction)\| + \varepsilon}{\|\velocity_{\rm{obs}}\| + \varepsilon} \right)\right)^2 dS 
    + 8\times 10^{-6} \int_{\Gamma_b} \|\nabla \friction\|^2 dS.
\end{aligned}
\end{equation}
Here, $\velocity(\friction)$ is the HO solution at $t=\tint$ for a given basal friction field $\friction$, $\Gamma_{\rm{s}}=\Gamma_{\rm{s}}(\tint)$ is the domain's initial surface boundary interpolated from \cite{howat2014greenland} (version 2), and $\velocity_{\rm{obs}}$ is the observed 1995-2015 averaged surface velocity field from \cite{MEaSUREs-Vel_v1}.
The variable $\varepsilon = \SI{2.22e-16}{m/s}$  is added to avoid division by zero.
The weights for the first two cost functions are normalization parameters chosen to balance the influence of the two terms; the weight for the regularization was chosen via an L-curve analysis.

The HO system
\eqref{hoxx} and \eqref{hoyy} 
is solved for the 3-dimensional velocity fields $v_x$ and $v_y$.
After they have been determined, the vertical velocity field $v_z$ can be constructed by from the constraint $\text{div} \velocity = 0$, which is required by the conservation of mass, assuming that ice is an incompressible material.
Note that although both the FS and HO models are posed on the same three-dimensional domain $\Omega(t)$, the HO model involves solving for only two unknowns ($v_x$ and $v_y$) instead of four ($v_x$, $v_y$, $v_z$, and pressure $p$).
Moreover, without the additional stability requirements posed by the pressure term \cite{chen2013well, leng2014finite}, the HO system is simpler to solve numerically.
Hence, the computational costs associated with the HO model are smaller than that for the FS model. 
For a detailed introduction and comparison of both equations, we refer to \cite{greve2009dynamics}.

\subsubsection{Thermal model}\label{sec:temperature}

The thermal model is derived from the balance equation of internal energy and Fourier's law of heat transfer under the assumption of constant heat conductivity $\kappa_{\rm{ice}} = \SI{2.4}{WK/m}$, constant heat capacity $c_{\rm{ice}} = \SI{2093}{JK/kg}$, and constant ice density $\rho_{\rm{ice}} = \SI{917}{kg/m^3}$. 
Given the velocity vector $\velocity(t) = (v_x(t), v_y(t), v_z(t))^{\top}$ at time $\tint \le t \le \tfin$, the thermal model is described by the PDE 
\begin{equation}
    \begin{aligned}\label{eq:thermal:PDE}
    &\rho_{\rm{ice}} c_{\rm{ice}} \left(
    \frac{\partial \temperature}{\partial t} + \velocity \cdot \nabla \temperature \right)
    = -  \kappa_{\rm{ice}} c_{\rm{ice}} \Delta \temperature + \text{tr}(\sigma \cdot \dot{\varepsilon})
\end{aligned}
\end{equation}
for $\mathbf{x} \in \Omega(t)$ and $\tint < t \le \tfin$, and
where $\sigma$ is the Cauchy stress tensor, $\dot{\varepsilon}$ the strain rate tensor, and $\text{tr}$ denotes the trace operator.
The PDE \eqref{eq:thermal:PDE} accounts for the transfer of energy (due to dissipation via the Fourier heat law), convection (due to ice movement), friction and viscous heating (due to ice deformation and sliding).

At the surfaces $\Gamma_{\rm{air}}$ of the ice sheet exposed to air, i.e., the upper surface and the part of the lateral surface above sea level, we locally prescribe the mean annual air temperature $\temperature_{\rm{air}}(\mathbf{x})$ predicted by CNRM-CM6-1 \cite{voldoire2018cnrm} as Dirichlet boundary condition.
As the geometry $\Omega(t)$ changes, $\temperature_{\rm{air}}$ is adjusted to account for changes in altitude as described in \cite{nowicki2020experimental}.
To preserve continuity of $\temperature$ between years, at the beginning of each year we linearly interpolate between the old and new air temperature over a duration of 9.125 days.
At the basal boundary $\Gamma_{\rm{b}}$ of the ice sheet touching the bedrock, we prescribe the Neumann boundary condition
\begin{equation*}
    \begin{aligned}
    &- \kappa_{\rm{ice}} c_{\rm{ice}}  \nabla \temperature \cdot \mathbf{n} = \heatflux(x,y)  
\end{aligned}
\end{equation*}
for $\mathbf{x}=(x,y,z)\in\Gamma_{b}$ and $\tint\le t\le \tfin$, and
where $\heatflux$ denotes the geothermal heat flux. 
Since the latter is only sparsely observable through ice core measurements, we model it as a random variable (see Section \ref{sec:uncertainties});
for our reference parameterization, we choose $\heatflux$ as the average of the fields in \cite{shapiro2004inferring} and \cite{greve2019geothermal}.

\subsubsection{Ice thickness equation}\label{sec:thickness}

The mass and geometry of the ice sheet are determined by the ice thickness $\height$, whose evolution is governed by the PDE
\begin{equation}\label{heighteq}
    \frac{\partial \height}{\partial t} = - \nabla \cdot (\height \bar{\velocity}) + \dot{M}_{\rm{s}} - \dot{M}_{\rm{b}}
\end{equation}
for $(x,y)\in\Omega_{\rm{2D}}$, and $\tint\le t\le \tfin$.
The PDE \eqref{heighteq} is connected to the velocity $\velocity$ of the ice sheet through the depth-averaged horizontal velocity vector $\bar{\velocity}$.
In addition, \eqref{heighteq} depends the imposed climate forcing and the ice temperature $\temperature$ through the basal melt rate $\dot{M}_{\rm{b}}$ and the surface mass balance $\dot{M}_{\rm{s}}$.
The basal melt $\dot{M}_{\rm{b}}$ is caused by ice-movement induced friction and geothermal heating; 
it is updated throughout the computations using the geothermal heat flux $\heatflux$, the ice velocity $\velocity$, and the ice temperature $\temperature$.
The surface mass balance $\dot{M}_{\rm{s}}$ is an input function that is positive wherever accumulation occurs, e.g., from snowfall. 
For our purposes here, we impose the surface mass balance predicted by the CNRM-CM6-1 climate model \cite{voldoire2018cnrm} with the altitude dependent adjustment described in \cite{nowicki2020experimental}.

Our OoI is the ice mass loss in the modelled domain of Greenland compared to a control run over the 2015--2050 time period.
The purpose of the control run is to negate model drift, and is computed using the CNRM-CM6-1 \cite{voldoire2018cnrm} SSP1-2.6 1960--1989 reference forcing, i.e., the same forcing as used during the initialization at equilibrium.
The ice mass is computed by integrating the ice thickness $\height$ over $\Omega_{\rm{2D}}$, and multiplying with the ice density $\rho_{\rm{ice}} = 
\SI{917}{kg/m^3}$.
Consequently, ice mass can be lost not only through melting, but also through ice discharge as ice is advected outside of the modelled domain along the periphery of the ice sheet.

\subsection{Model Uncertainties}\label{sec:uncertainties}

Ice sheet simulations are subject to a number of uncertainties because many of the physical properties cannot be directly observed through measurements and their indirect inference from data leads to model parameter uncertainties.
Moreover, projections are further uncertain because future forcing conditions are unknown.
Reports to inform policy making treat this issue by defining socioeconomic pathway scenarios that climate models can simulate.
Ice sheet simulations use these climate model projections to define the future forcing, while parameter uncertainties are treated by comparing projections of models from different expert groups with each other for the same scenarios (e.g., ISMIP6 \cite{ismip6, seroussi2020ismip6}).
However, these approaches do not result in formal forward propagation of parameter uncertainties to determine quantitative effects on projections; rather, the ensemble approach computes a mean projection by averaging across the projections of different models (typically a small number). We address that gap here through the introduction of UQ methods that formally characterize parameter uncertainties and compute the impact of those uncertainties on projections.

\begin{figure*}
    \centering    
    \includegraphics[width=\textwidth, trim=160 190 100 160, clip]{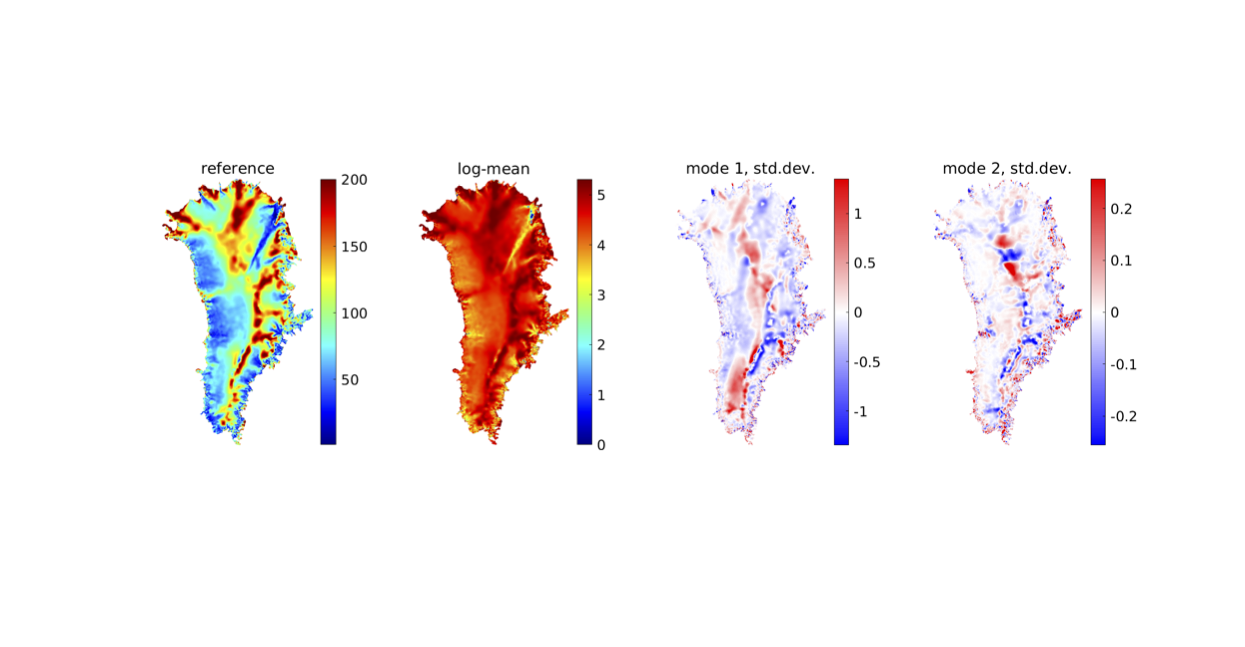}
    \caption{
    Far left: Reference basal friction field at the parametric mean (measured in $\mathrm{(s / m)^{1/2}}$);
    Center left to right:
    mean and first two modes (scaled by their standard deviation) of the log-normal distribution (measured in $\mathrm{\ln(s / m)/2}$)
    }
    \label{fig:basalfriction}
\end{figure*}

We first characterize uncertainty in the basal friction field $\friction$, which governs the sliding of the ice at the bedrock and thus has a strong influence on the ice velocity $\velocity$.
However, the basal friction field can only be indirectly observed through velocity data at the surface, and is thus uncertain.
Estimates of the basal friction field are commonly obtained by solving an inverse problem such as \eqref{eq:hf:costfunction}. 
We model the uncertainty in the basal friction field induced by the choice for $\velocity_{\rm{obs}}$:
Using the same weights as in \eqref{eq:hf:costfunction}, we compute  basal friction fields $\friction_{i}$ as the minimizers of the cost function $J(\alpha; \velocity^{\rm{obs}}_i)$ with $\velocity^{\rm{obs}}_i$ being the annual average surface velocity data for the year $i$ ($i=2015, \dots, 2021$) from \cite{MEaSUREs-Annual}.
In these inverse problems we use the geometry of our 2015 initial condition, and start the minimization from $\friction_{\rm{ref}}$ as initial guess, such that the data $\velocity^{\rm{obs}}_i$ can loosely be interpreted as measurements obtained for a basal friction $\friction$ drawn from a log-normal distribution
\begin{align}\label{eq:friction}
    \friction = \exp\left(\ln(\friction_{ref}) + \sum_{i=1}^7 X_{\alpha, i} \phi_i \right),
\end{align}
with random variable $X_{\alpha} \in \mathbb{R}^{7}$, $X_{\alpha} \sim \mathcal{N}(0, \Sigma_{\alpha})$, and modes $\phi_1, \dots, \phi_7$ defined on the basal surface $\Gamma_{\rm{b}}$.
Note that by modeling $\friction$ through a log-normal distribution, we are guaranteed that $\friction$ is positive everywhere.
To determine $\Sigma_{\alpha} \in \mathbb{R}^{7 \times 7}$ and $\phi_1, \dots, \phi_7$, we form the covariance matrix between $\ln(\friction_{i}) - \ln(\friction_{ref})$, $i=2015, \dots, 2021$; we then choose $\phi_i$ as the $i$-th principal component and $\Sigma_{\alpha}$ as the associated diagonal covariance matrix (c.f., \cite{abdi2010principal}).
Figure \ref{fig:basalfriction} shows the log-mean and the two dominant principal components 
scaled by their respective standard deviation.
The first mode identifies large areas with strong uncertainty, in the second the uncertainties are smaller and slightly more local, but still affect large areas of the domain.
The remaining modes (not shown) follow this trend in reflecting increasingly more local fluctuations in the basal friction field $\friction$.

\begin{figure}
    \centering
    \includegraphics[width=0.5\linewidth, trim=100 20 50 20, clip]{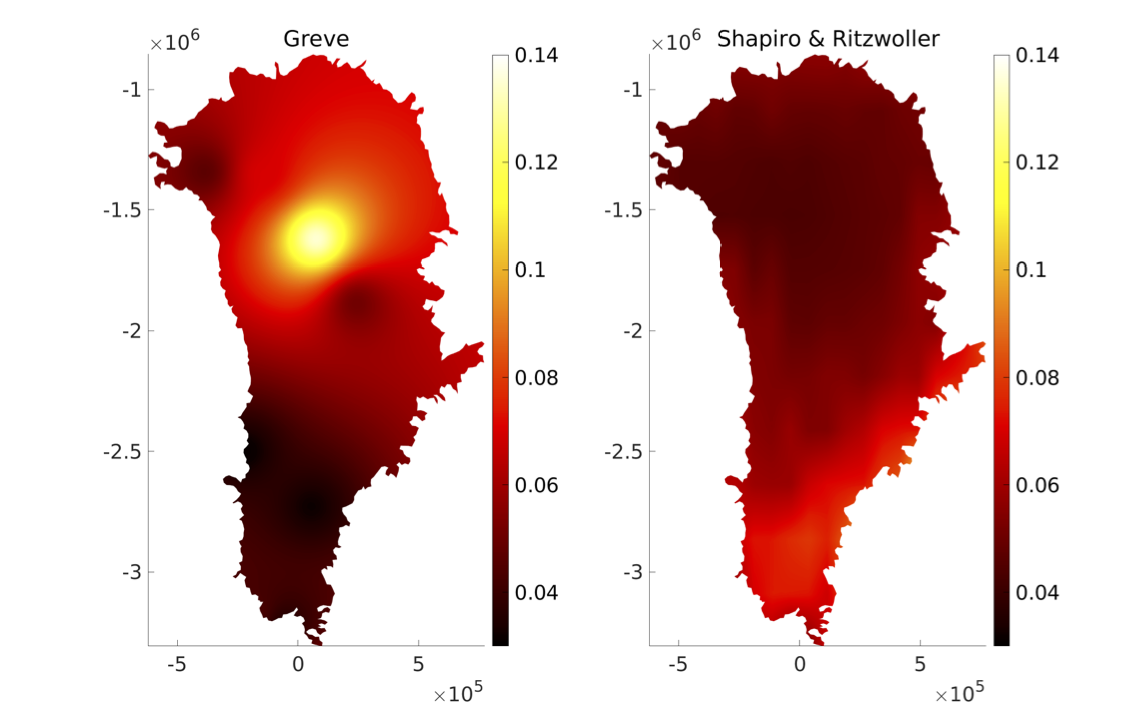}
    \caption{
    Geothermal heat flux fields from \cite{greve2019geothermal} (left) and \cite{shapiro2004inferring} (right) measured in \SI{}{W/m^2}
    }
    \label{fig:heatflux}
\end{figure}

Next, we characterize the uncertainty in the geothermal heat flux, which describes how much heat moves outward from the interior of the earth and is needed for the basal boundary condition of the thermal model.
Since the basal boundary is underneath the ice, the geothermal heat flux is difficult to observe and therefore has large uncertainty \cite{rogozhina2012effects}.
Sparse measurements are obtained through ice cores.
There are many approaches using these sparse geothermal heat flux measurements to infer the complete geothermal heat flux field, including machine learning \cite{colgan2021greenland, rezvanbehbahani2017predicting}, topographic corrections \cite{colgan2021topographic}, magnetic fields \cite{martos2018geothermal, kolster2023satellite}, and others \cite{artemieva2019lithosphere, rogozhina2012effects}.
Different approaches lead to flux fields that are vastly different (see \cite{colgan2021greenland}, Figure 13, or \cite{rogozhina2012effects}), and there generally appears to be disagreement on how to best quantitatively describe the flux field.
For instance, in ISMIP6, 13 groups used the heat flux from \cite{shapiro2004inferring}, where seismic data is used to extrapolate heat flow measurements, five groups used \cite{greve2019geothermal}, where ice-core measurements are used for local adjustments of the global heat flux map \cite{pollack1993heat}, and the remaining groups used different fluxes (\cite{ismip6}, Table A1).
Yet, as shown in Figure~\ref{fig:heatflux}, the geothermal heat flux fields from \cite{greve2019geothermal} and \cite{shapiro2004inferring} are considerably different.

We build an uncertainty model for the  geothermal heat flux using the fields that were predominantly chosen in the ISMIP6 models: the field $\heatfluxG$ with superscript  ``G'' for ``Greve" from \cite{greve2019geothermal}, and the field $\heatfluxSR$ with superscript ``SR'' for ``Shapiro-Ritzwoller" from \cite{shapiro2004inferring}, both depicted in Figure~\ref{fig:heatflux}.
Following the example at the NASA Sea Level Change Portal,\footnote{https://sealevel.nasa.gov/, accessed Aug.~2024} we model the uncertainty of the geothermal heat flux through
\begin{align*}
    \heatflux = X_{\rm{geo}} \heatfluxG + (1-X_{\rm{geo}}) \heatfluxSR 
\end{align*}
where $X_{\rm{geo}} \sim \mathcal{U}(0,1)$ is a uniformly distributed random variable taking values between 0 and 1.

\begin{figure*}
    \centering
    \includegraphics[width=\textwidth]{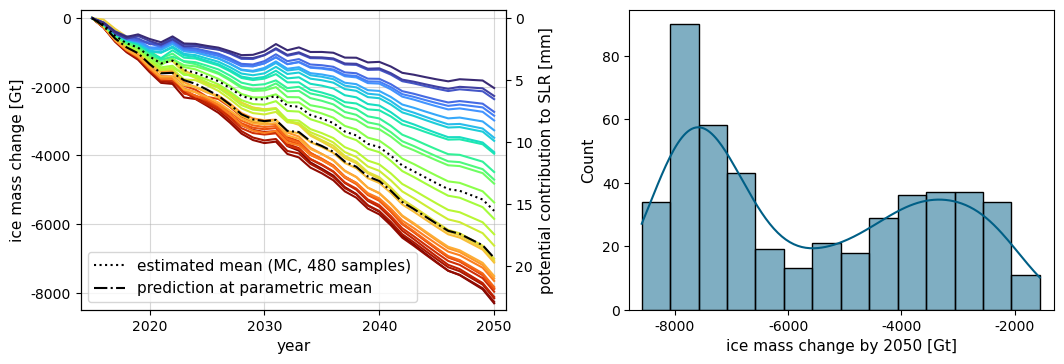}
    \caption{
    Left: 
    High-fidelity ice mass change projections for 32 samples (colored lines) of the geothermal heat flux and basal friction fields;
    right: 
    Histogram and kernel density estimate of the high-fidelity ice mass change projections for the year 2050 computed over 480 samples of the geothermal heat flux and basal friction fields
    }
    \label{fig:predictions}
\end{figure*}

We now show quantitatively the effects of these uncertainties in basal friction field and geothermal heat flux on predicted ice mass loss, treating $X_{\rm{geo}}$ and $X_{\alpha}$ as independent variables.
This choice neglects how the inversion for the basal friction field would absorb increased/decreased ice velocities caused by stronger/weaker geothermal heating (see \cite{issm}), and therefore leads to higher/lower local ice velocity.
To additionally model this effect, the uncertainties in $\heatflux$ need already be included when generating the training data for the model \eqref{eq:friction} such that $(X_{\rm{geo}}, X_{\alpha})$ can be modelled as a correlated random vector.
Here, we used the mean $\mathbb{E}(\heatflux)$ when building the model \eqref{eq:friction}.

Figure~\ref{fig:predictions} shows the high-fidelity projections of the 2015--2050 ice mass loss in Greenland for 32 samples of $(X_{\rm{geo}}, X_{\alpha})$ chosen via Latin-Hypercube sampling.
The colors are ordered according to the magnitude of the projections for the year 2050.
It can be seen that the modelled uncertainty in the geothermal heat flux and the basal friction field lead to large variations in the predicted ice mass change. 
These variations grow increasingly as time progresses.
Comparing the mean ice mass change over 480 projections of i.i.d.\ samples (dotted line) with the predicted ice mass change at the mean parameter $\mathbb{E}[(X_{\rm{geo}}, X_{\alpha})] = (0.5, 0, \dots, 0)$ (dashed-dotted line), we see a discrepancy of \SI{1310}{Gt} by the year 2050.
This discrepancy is perhaps not surprising considering that our high-fidelity model is highly nonlinear with respect to the uncertain parameters.  
Moreover, the histogram of 480 samples plotted in Figure~\ref{fig:predictions} illustrates that the model nonlinearity leads to a non-Gaussian distribution in ice mass change outputs, likely due to the presence of tipping points in ice retreat.
This highlights the critical importance of conducting a full quantitative assessment of uncertainty.

We remark that the multifidelity UQ framework introduced in Section \ref{sec:methods} is agnostic to the specific choice of modeled uncertainties; thus, other characterizations of the uncertainty in the geothermal heat flux and basal friction fields are permitted, e.g., through Gauss Markov random fields (see \cite{bulthuis2022implementation}), posterior distributions (see \cite{isaac2015scalable, babaniyi2021inferring}), or learned distributions (see \cite{brinkerhoff2021constraining, rezvanbehbahani2017predicting}).
Similarly, other uncertainties may be considered, such as the basal topography (see \cite{morlighem2017bedmachine, lampkin2011preliminary}), or hyperparameters associated with data-fit surrogate models.

\subsection{Surrogate modeling}\label{sec:surrogates}

The multifidelity UQ framework relies on the availability of surrogate lower-fidelity/lower-cost models.
In this section we provide brief descriptions of some of the types of surrogate models readily available in many ice sheet codes.

A straightforward means for obtaining lower-fidelity/lower-cost surrogates is to use coarser grids and/or larger time steps when discretizing the continuous models. This approach was first used for MLMC estimation in \cite{heinrich2001multilevel} and further developed in \cite{giles2008multilevel}. Because a coarsened computational model has fewer degrees of freedom, its solution can be expected to require less computational effort. Note that when coarsening a discretization, it is important to still adhere to stability conditions (e.g., the CFL condition for time step size).

Another common form of surrogate model in computational glaciology is a simplified-physics model, especially for the stress balance equations.
The development of simplified-physics approximations is primarily motivated by the high computational cost of the FS equations, which necessitates surrogate modeling for many applications.
Consequently, there exists a vast variety of approximated physics models, primarily the HO equations (if the high-fidelity model is the Full Stokes system), SSA equations \cite{MacAyeal1989}, the Shallow Ice Approximation \cite{le2004glacier}, mono-layer models \cite{brinkerhoff2015dynamics,dias2022new}, and hybrid models that impose different equations on different parts of the domain \cite{pattyn2010antarctic,pollard2012description}.
The prevalence of approximated physics models in the ISMIP6 studies \cite{ismip6,goelzer2018design, seroussi2020ismip6, seroussi2019initmip} and throughout the ice sheet literature shows the existing trust in these methods.
It can thus be expected that approximated physics models have strong correlations with the high-fidelity model, and are already implemented in many ice sheet codes.

In the SSA model, vertical shear is neglected such that only the depth-averaged velocity field $\bar{\velocity}$ on the 2D domain $\Omega_{\rm{2D}}$ is solved for.
The governing equations become:
\begin{align}
    \nabla \cdot (2\bar{\mu} \height \strainrate_{\rm{SSA}, 1}) - \friction^2 N\bar{\velocity}_x &= \rho g \height \ddx{s}\label{eq:SSA:x} \\
    \nabla \cdot (2\bar{\mu} \height \strainrate_{\rm{SSA}, 2}) - \friction^2 N\bar{\velocity}_y &= \rho g \height \ddy{s} \label{eq:SSA:y}
\end{align}
for $\mathbf{x}=(x,y) \in \Omega_{\rm{2D}}$ and $\tint < t \le \tfin$, and
where the mean ice viscosity $\bar{\mu}$ is computed from Glen's flow law \eqref{eq:Glen} with strain rate
\begin{align*}
    \strainrate_{\rm{SSA}, 1} = \left(
    \begin{array}{c}
        \displaystyle 2 \ddx{\bar{\velocity}_x} + \ddy{\bar{\velocity}_y} \\
        \displaystyle \frac12 \ddy{\bar{\velocity}_x} + \frac12 \ddx{\bar{\velocity}_y}
    \end{array}
    \right)
\end{align*}
and
\begin{align*}
    \strainrate_{\rm{SSA}, 2} = \left(
    \begin{array}{c}
        \displaystyle \frac12 \ddy{\bar{\velocity}_x} + \frac12 \ddx{\bar{\velocity}_y} \\
        \displaystyle \ddx{\bar{\velocity}_x} + 2\ddy{\bar{\velocity}_y}
    \end{array}
    \right).
\end{align*}
The equations \eqref{eq:SSA:x}, \eqref{eq:SSA:y} are 
solved with boundary conditions
\begin{align*}
    \strainrate_{\rm{SSA}, 1} \cdot \mathbf{n}_{\rm{2D}} &= 0 
    & \strainrate_{\rm{SSA}, 2} \cdot \mathbf{n}_{\rm{2D}} &= 0 
    & \text{on } \partial \Omega_{\rm{2D}},
\end{align*}
on the interface of $\Omega_{\rm{2D}}$ with air, and with
\begin{align*}
    2\bar{\mu} \height \strainrate_{\rm{SSA}, 1} \cdot \mathbf{n}_{\rm{2D}} &= f_{
    \rm{w, SSA}} n_x \\
    2\bar{\mu} \height \strainrate_{\rm{SSA}, 2} \cdot \mathbf{n}_{\rm{2D}} &= f_{\rm{w, SSA}} n_x
\end{align*}
on the ice-water interface.
Here, $\mathbf{n}_{\rm{2D}} = (n_x, n_y)^{\top}$ denotes the outward pointing unit normal of the domain $\Omega_{\rm{2D}}$.
The water pressure is applied through the Neumann flux function
\begin{align*}
    f_{\rm{w, SSA}} = \frac12 ( \rho_{\rm{ice}} g \height^2 - \rho_{\rm{w}} g b^2).
\end{align*}
Since 
\eqref{eq:SSA:x}, \eqref{eq:SSA:y} are 
solved for two variables defined over a 2D surface instead of two variables defined over a 3D domain, SSA models offer vast computational savings compared to HO models:
For the meshes considered in this study, predictions using an SSA model were between 142 and 1,099 times faster than their HO counterparts, see Table \ref{tab:discretization}.

\begin{table*}[t]
    \centering
    \begin{tabular}{@{}l|rr|rr|rr@{}}
    \toprule
        & \multicolumn{2}{c|}{resolution} & \multicolumn{2}{c|}{DoFs/variable} & \multicolumn{2}{c}{cost [CPUh]} \\
        mesh name & min. & max. & 2D & 3D & SSA & HO \\
        \midrule
        fine & 100 m & 15 km & 20,455 & 102,275 & 0.455 & 64.888 \\
        medium & 100 m & 30 km & 6,554 & 32,770 & 0.105 & 22.889 \\
        coarse & 100 m & 50 km & 3,600 & 18,000 & 0.059 & 4.689 \\
        \botrule
    \end{tabular}
    \caption{    
    Discretization and runtime of our ISSM models
    }
    \label{tab:discretization}
\end{table*}

Another tool for building surrogate models is training on available data of the high-fidelity model or its approximations.
Most prevalent are emulators of the model's OoIs, for instance through Gaussian process regression;
this approach was taken in the forward UQ studies \cite{little2013upper, bulthuis2019uncertainty, pollard2016large, aschwanden2019contribution, edwards2021projected}.
Other forms of data-driven models include 
physics-based learning keeping the connection to governing equations 
\cite{riel2023variational} and more general machine learning approaches \cite{jouvet2022deep,he2023hybrid}.
Throughout these approaches, training-based surrogate models hinge on the availability of training data, require some amount of expertise from the modeler, and can have large generalisation errors outside the training regime, especially if the uncertain parameters are high-dimensional.
On the other hand, they can achieve significantly larger speed-ups than physical approximations.

\begin{figure*}
    \centering
\includegraphics[width=\textwidth]{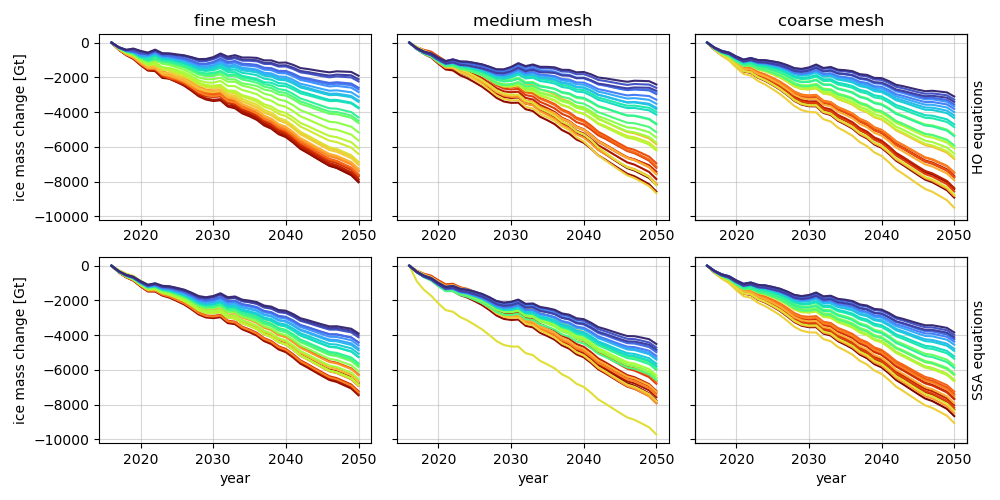}
    \caption{Effect of mesh coarsening (left to right) and physical approximation (top to bottom) on the predicted 2015-2050 ice mass change for 32 samples of the geothermal heat flux and basal friction fields relative to control run on the same mesh with the same physical approximation
    }
    \label{fig:predictions:2x3}
\end{figure*}

In general, for multifidelity UQ any approximation of the high-fidelity input/output map can be considered as a surrogate model as long as we can characterize its cost and its correlation with the high-fidelity model. However, despite the computational gains offered by surrogate models, the high-fidelity model should generally not be replaced when estimating its OoI as the model bias introduced by the surrogate can be significant.
For example, Figure~\ref{fig:predictions:2x3} illustrates the effects of both mesh coarsening and physical approximation via SSA on ice mass change predictions for 32 samples of the geothermal heat flux and the basal friction fields chosen by Latin Hypercube sampling, with colors identifying predictions for the same samples in different plots.
Considering the fine mesh with the HO equations to be the highest fidelity model, for the five surrogate models in Figure~\ref{fig:predictions:2x3}, the mean 2050 ice mass change differs by \SI{407}{Gt} to \SI{2286}{Gt} (\SI{13.7}{\%} to \SI{77.0}{\%}) from that estimated from the high-fidelity samples. 
To avoid such an introduction of model bias, multifidelity estimators do \textit{not} replace the high-fidelity model with surrogates, but incorporate multiple models into their structure in a way that guarantees an \textit{unbiased} estimate of the statistics of the high-fidelity OoI.

\section{Multifidelity Uncertainty Quantification}\label{sec:methods}

This section presents multifidelity UQ methods that incur lower costs than Monte Carlo estimation by exploiting the statistical correlation between the high-fidelity model and its surrogates.
We present preliminaries (Section~\ref{sec:preliminaries}) and
then discuss three different methods
with complementary strengths that make them appropriate in different settings:
\begin{itemize}
    \item MFMC (Section~\ref{sec:mfmc}): Handles a large variety of model types and computational costs, e.g., achieved by physical approximations and data-driven surrogates, including situations where there is no clear a priori model hierarchy;
    \item MLMC (Section~\ref{sec:mlmc}): Is particularly appropriate when there is a hierarchy of similar lower-fidelity models, e.g., achieved by mesh coarsening;
    \item MLBLUE (Section~\ref{sec:mlblue}): 
    By construction is always at least as good as MLMC and MFMC, but implementation and handling can be more involved.
\end{itemize}
We also note that other multifidelity UQ methods, such as the Approximate Control Variate approach \cite{gorodetsky2020generalized}, introduce alternative algorithmic formulations that may be beneficial in some applications settings. 
For MFMC, MLMC, and MLBLUE we provide an algorithmic description for choosing an optimized subset of surrogate models with optimal weights and sample sizes, and the computation of the associated multifidelity estimator.
The sample sizes are chosen such that the estimators' total costs remain within a given computational budget $c>0$.\footnote{We note that each algorithm can be re-written to work with a target accuracy instead.} 
The differences between the methods and their applicability for UQ in ice sheet simulations are discussed in Section \ref{sec:discussion}.

\subsection{Preliminaries}
\label{sec:preliminaries}

In the following, we denote with $s_1$ a high-fidelity output of interest (OoI), and by $s_1(\theta) \in \mathbb{R}$ its scalar\footnote{
We restrict our exposition to scalar OoIs solely for the purpose of simpler notation; each of the presented multifidelity UQ methods can be posed similarly for vector-valued quantities.
For MLBLUE, it is even possible to consider surrogate models that have additional or less OoIs than the high-fidelity model (c.f., \cite{Croci2023}).
} evaluation at a parameter sample $\theta \sim \nu$. 
In addition, we have at hand $L-1 \in \mathbb{N}$ surrogate OoIs $s_2, \dots, s_L$.
Evaluating an output $s_i$ for a parameter $\theta \sim \nu$ incurs the computational cost $c_i > 0$, e.g., measured in CPU-hours.
We assume without loss of generality that the models are ordered by cost: $c_1 \ge c_2 \ge \dots \ge c_L$.
In our setting here, $s_1$ is the 2050 ice mass change (in Gt) computed by the high-fidelity model described in Section \ref{sec:high-fidelity}, and $\nu = \mathcal{U}(0, 1) \times \mathcal{N}(\mathbf{0}, \Sigma)$ is the probability distribution of the uncertainties in the geothermal heat flux and basal friction field described in Section \ref{sec:uncertainties}.

Furthermore, for $i,j = 1,\ldots,L$, we define the model covariance matrix $\Sigma \in \mathbb{R}^{L \times L}$:
\begin{align*}
    \Sigma_{i,j} &:= \text{$\mathbb{C}$ov}(s_i, s_j) \\
    &\hphantom{:}= \mathbb{E}_{\nu}[(s_i - \mathbb{E}_{\nu}[s_i])(s_j - \mathbb{E}_{\nu}[s_j])].
\end{align*}
If this matrix is not available from theoretical estimates, it can be approximated using model convergence rates or from $n_{\rm{pilot}}$ pilot samples:
\begin{align*}
    \Sigma_{i,j} \hspace{-0.1em}\approx\hspace{-0.1em} \frac{1}{n_{\rm{pilot}}\hspace{-0.15em}-\hspace{-0.15em}1}\hspace{-0.15em} \sum_{k=1}^{n_{\rm{pilot}}} \hspace{-0.1em}(s_i(\theta_k) \hspace{-0.1em}-\hspace{-0.1em} \bar{s}_i^{\rm{pilot}})\hspace{-0.1em}
    (s_j(\theta_k) \hspace{-0.1em}-\hspace{-0.1em} \bar{s}_j^{\rm{pilot}})
\end{align*}
with $\bar{s}_i^{\rm{pilot}} := n_{\rm{pilot}}^{-1} \sum_{k=1}^{n_{\rm{pilot}}} s_i(\theta_k)$.
We expect that for most ice sheet applications, sufficient data is available from validation and verification procedures, as well as from historic, control, and spin-up runs. 
However, should additional sampling indeed be necessary, the model evaluations may be re-used when evaluating the multifidelity estimators.
We therefore do not count the estimation of $\Sigma$ towards computational costs.

\subsection{Multifidelity Monte Carlo}\label{sec:mfmc}

The MFMC method employs surrogate models of arbitrary structure \cite{ng2014multifidelity,Peherstorfer2016b}.
In the following we focus on the MFMC method from \cite{Peherstorfer2016b} with the adaptation for restrictive computational budgets from \cite{Gruber2023} and an integrated model selection step.
The method is summarized in Algorithm \ref{alg:mfmc}.
We note that MFMC has been extended to 
estimation of sensitivities and covariances \cite{qian2018multifidelity},
is compatible with goal-oriented training of surrogate models \cite{peherstorfer2019multifidelity, farcaș2023context}, and has been applied to large-scale applications, 
in particular in climate modeling \cite{gruber2023climate}.

The MFMC estimator is based on writing the target statistic $\mathbb{E}[s_1]$ in the form
\begin{samepage}
\begin{align*}
    \mathbb{E}[s_{1}] 
    &= \mathbb{E}[s_{1}] + \alpha_{2} \mathbb{E}[s_{2}] - \alpha_{2} \mathbb{E}[s_{2}] \\
    &= \mathbb{E}[s_{1}] + \sum_{j = 2} ^{L} \alpha_{j} \big(\mathbb{E}[s_{j}] - \mathbb{E}[s_{j}] \big)
\end{align*}
\end{samepage}
for arbitrary weights $\alpha_{2}, \dots, \alpha_{L}$ (optimal values are defined below).
Each expectation $\mathbb{E}[s_{j}]$ is then approximated using the Monte Carlo method with shared samples such that the MFMC estimator $\bar{s}_{\rm{MFMC}}$ has the form
\begin{equation}\label{mfmcest}
    \begin{aligned}
    &\bar{s}_{\rm{MFMC}}
        :=
    \frac{1}{n_1} \sum_{i=1}^{n_1} s_{1}(\theta_{i}) \\
    &\quad + \sum_{j = 2}^{L} \alpha_{j} \left( 
    \frac{1}{n_{j}} \sum_{i=1}^{n_{j}} s_{j}(\theta_{i})
    - \frac{1}{n_{j-1}} \sum_{i=1}^{n_{j-1}} s_{j}(\theta_{i})
    \right)
\end{aligned}
\end{equation}
where, for each $1\le j \le L$, $n_j \in \mathbb{N}$ is the number of samples for which the OoI $s_j$ needs to be evaluated using model $j$, and $\theta_{1}, \dots, \theta_{n_{L}} \sim \nu$ are i.i.d. parameter samples shared between the models.
By imposing the order $1 \le n_1 \le n_{2} \le \dots \le n_{L}$, the cheapest model $s_L$ is sampled the most while the expensive high-fidelity model, $s_1$, is sampled the least.
By construction, the MFMC estimator $\bar{s}_{\rm{MFMC}}$ is unbiased: $\mathbb{E}\big[\bar{s}_{\rm{MFMC}}\big]=\mathbb{E}[s_{1}]$.
Note that, in constrast to MLMC, the samples within the individual sums in \eqref{mfmcest} are shared.
Exploiting the consequent correlation between the sums, the mean squared error (MSE) of the MFMC estimator \eqref{mfmcest} is then given by
\begin{equation}\label{mfmcvar}
\begin{aligned}
&\text{MSE}(\bar{s}_{\rm{MFMC}})\\
&= \mathbb{E}[(\bar{s}_{\rm{MFMC}}-\mathbb{E}[s_1])^2]\\
&=  
\frac{\sigma_1^2}{n_1} + \sum_{j=2}^{L} \left(\frac{1}{n_{j-1}} - \frac{1}{n_{j}} \right) (\alpha_{j}^2\sigma_{j}^2 - 2\alpha_{j} \correlation{j} \sigma_{1} \sigma_{j})
\end{aligned}
\end{equation}
where, for $1 \le j \le L$, $\sigma_j = \sqrt{\Sigma_{j,j}}$ is the standard deviation for model $s_j$, and $\correlation{j} := \Sigma_{1,j}/(\sigma_1 \sigma_j)$ is the correlation of model $j$ with the high-fidelity model $s_1$.
Since the MFMC error is unbiased, the MSE is the expected squared error between the MFMC estimator and the target statistic $\mathbb{E}[s_1]$.
From the formula \eqref{mfmcvar} we can immediately deduce that $\displaystyle \alpha_{j} ={ \correlation{j} \sigma_1}/{ \sigma_{j} } = \Sigma_{1,j} / \Sigma_{j,j}$ for $j=2, \ldots,L$ are optimal for achieving the smallest MSE in \eqref{mfmcest}.

To obtain optimal sample sizes $1 \le n_1 \le n_{2} \le \dots \le n_{L}$ for a given total computational budget $c \ge c_{1}$, we solve the relaxed optimization problem
\begin{equation}
\begin{aligned}\label{eq:MFMC:samplesizeminimization}
&\min_{n_1, \dots, n_{L} \in \mathbb{R}} \text{MSE}(\bar{s}_{\rm{MFMC}}) \\
&\quad \text{ s.t. } \quad \left\{
\begin{array}{l}
0 < n_1 \le n_{2} \le \dots \le n_{L} \\
\sum_{j=1}^{L} n_j c_j \leq c.
\end{array}
\right.
\end{aligned}
\end{equation}
The first constraints on the sample sizes ensure that at least as many samples are taken of the OoI $s_{j}$ as are taken for the more costly OoI $s_{j-1}$, with the constraint $0 < n_1$ ensuring that all sample sizes are positive.
The last constraint on the model evaluation costs ensures that the given computational budget $c$ is not exceeded.

It has been shown in \cite{Peherstorfer2016b} that if $1 = |\correlation{1}| > |\correlation{2}| > \dots > |\correlation{L}|$ and
\begin{align}\label{eq:MFMC:requirement}
    \frac{c_{j-1}}{c_j} > \frac{\correlation{j-1}^2 - \correlation{j}^2}{\correlation{j}^2 - \correlation{j+1}^2}
\end{align}
hold, then the global minimum of \eqref{eq:MFMC:samplesizeminimization} is obtained by
\begin{align}\label{eq:MFMC:closedform}
    n_1 = c \left(\sum_{j=1}^{L} c_j r_{j} \right)^{-1}, \quad
    n_j = r_{j} n_1 
\end{align}
for $j = 2, \dots, L$,
where we are using the auxiliary variable 
\begin{align}\label{eq:MFMC:ratio}
    r_{j} := \sqrt{\frac{c_{1}(\correlation{j}^2-\correlation{j+1}^2)}{c_{j}(\correlation{1}^2-\correlation{2}^2)}}.
\end{align}
Condition \eqref{eq:MFMC:requirement} describes how much faster any lower-fidelity model $s_j$ must be, compared to the more expensive model $s_{j-1}$, in order to make up for its worse correlation $|\correlation{j}| < |\correlation{j-1}|$.\footnote{Note that if there exists a model $j+1$ such that $\correlation{j+1} \ge \correlation{j}$, then it is clearly superior to model $s_j$ both in cost and correlation, and model $s_j$ should be removed.}
After computing $n_1, \dots, n_L \in \mathbb{R}$ using the closed-form solution \eqref{eq:MFMC:closedform}, they need to be rounded to integers larger or equal than one.
If possible, sample sizes are rounded down to ensure the computational budget is not exceeded.
Any remaining budget can be redistributed to further decrease the MFMC estimator's variance, e.g., using the optimized rounding strategy in \cite{Gruber2023}.

\begin{algorithm}[t]
\textbf{Input:  }{high-fidelity model $s_1$, surrogate models $s_2, \dots, s_L$, model evaluation costs $c_1 \ge \dots \ge c_L$, model correlations $\correlation{1}, \dots, \correlation{L}$, model variances $\sigma_1^2, \dots, \sigma_L^2$, computational budget $c$}\\
\textbf{Output: }{sample sizes $1 \le n_1 \le \dots \le n_L$, optimal weights $\alpha_1, \dots, \alpha_L$, estimator MSE $\texttt{MSE} = \mathbb{E}[(\bar{s}_{\rm{MFMC}}-\mathbb{E}[s_1])^2]$}
\begin{enumerate}
    \item Assert that $1=|\correlation{1}| > \dots > |\correlation{L}|$. If not, then any model $s_j$ with $\correlation{j} \le \correlation{j+1}$ needs to be removed; return $\mathtt{MSE}=\infty$.
    \item Define the auxiliary variable $\correlation{L+1} = 0$.
    \item Assert that the budget the budget is large enough, i.e., $\sum_{j=1}^{L} c_{j} \le c$, and that 
    \eqref{eq:MFMC:requirement}
    holds for $j = 2, \dots, L$. 
    If not, then the provided combination of surrogate models is not suitable for MFMC; return $\mathtt{MSE}=\infty$.
    \item Compute the ratios $r_j$ for $j=1, \dots, L$ using \eqref{eq:MFMC:ratio}, and use them to compute $n_1, \dots, n_L \in \mathbb{R}$ with \eqref{eq:MFMC:closedform}.
    \item For $j=1, \dots, L$, round all $n_j < 1$ up to 1, and round all $n_j \ge 1$ down to the next smallest integer.
    \item Compute the optimal weights $\alpha_j = \correlation{j}\sigma_1/\sigma_j$ for $j=1, \dots, L$
    \item Compute the MSE of the MFMC estimator with \eqref{mfmcvar} and save in variable \texttt{MSE}
    \item Return $(n_1, \dots, n_L)$, $(\alpha_1, \dots, \alpha_L)$, \texttt{MSE}
\end{enumerate}
\caption{Multifidelity Monte Carlo}
\label{alg:mfmc}
\end{algorithm}

Algorithm \ref{alg:mfmc} summarizes the computation of the sample sizes $n_1, \dots, n_L$ and weights $\alpha_1, \dots, \alpha_L$ for given models $s_1, \dots, s_L$.
Since the smallest MSE of the MFMC estimator may be achieved with a subset of the available surrogate models, Algorithm \ref{alg:mfmc} should always be placed within an outer loop over all subsets of the available surrogate models to determine when the MSE is the smallest.
In addition, the obtained value for $\mathtt{MSE}$ should be compared with the MSE of Monte Carlo sampling, i.e., the value $\sigma_1^2 / \lfloor c/c_1 \rfloor$, to ensure that MFMC is indeed applicable.
After the model subset has been selected, the MFMC estimator $\bar{s}_{\rm{MFMC}}$ can be computed using \eqref{mfmcest}.

\subsection{Multilevel Monte Carlo}\label{sec:mlmc}

The MLMC method has its roots in \cite{heinrich2001multilevel,giles2008multilevel}, where it was first formulated for grid-refinement based surrogate models and their convergence rates.
The method has since been expanded structurally (e.g., Multi-Index Monte Carlo \cite{haji2016multi}), refined for stronger convergence (e.g. randomized and adaptive MLMC methods \cite{mcleish2011general,blanchet2015unbiased,hoel2011adaptive}), and combined with other sampling schemes (e.g., Multilevel Quasi Monte Carlo \cite{giles2009quasi}, Multilevel Markov Chain Monte Carlo \cite{dodwell2019multilevel}); we refer to \cite{giles2015multilevel} for an extensive introduction.
In our exposition here we focus on the \textit{non-geometric} MLMC method, which is posed for arbitrary surrogate models and based on model correlations instead of convergence rates, following the exposition in \cite{giles2015multilevel}.
Algorithmic instructions are provided in Algorithm \ref{alg:mlmc}.

The MLMC estimator exploits that the expectation is a linear operator to expand the high-fidelity expectation $\mathbb{E}[s_1]$ in a telescoping sum of the form\footnote{
Contrary to typical MLMC notation but consistent with all other parts of this paper, a larger index here indicates a model of \textit{lower} computational cost, i.e. $c_L \le c_{L-1} \le \dots \le c_1$.
}
\begin{align*}
    \mathbb{E}[s_1]
    &= \mathbb{E}[s_2] + \mathbb{E}[s_1-s_2]\\
    &= \mathbb{E}[s_L] + \sum_{j=1}^{L-1} \mathbb{E}[s_{j}-s_{j+1}].
\end{align*}
The difference $s_{j}-s_{j+1}$ between models $s_{j}$ and $s_{j+1}$ is typically referred to as the $j$-th level, with $s_{L} = s_L-0$ the $L$-th level.
To obtain the MLMC estimator, the expectation of each level is approximated \textit{independently} via Monte Carlo sampling
\begin{equation}\label{eq:MLMC}
    \begin{aligned}
        &\bar{s}_{\rm{MLMC}} := \frac{1}{n_L} \sum_{i=1}^{n_L} s_L(\theta_i^{(L)}) \\
        &\quad + \sum_{j=1}^{L-1} \frac{1}{n_{j}} \sum_{i=1}^{n_{j}} \left(s_{j}(\theta_i^{(j)}) - s_{j+1}(\theta_i^{(j)}) \right) 
    \end{aligned}
\end{equation}
where, for each level $1 \le j \le L$, the $\theta_i^{(j)}\sim \nu$, $1 \le i \le n_{j}$, are i.i.d.\ samples, and $n_{j} \in \mathbb{N}$ is the sample size for the $j$-th level Monte Carlo approximation.
For implementation it is important to note that the samples are not shared beyond each level.
The cost of evaluating \eqref{eq:MLMC} is thus $n_L c_L + \sum_{j=1}^{L-1} n_{j} (c_{j} + c_{j+1})$.

By construction, $\bar{s}_{\rm{MLMC}}$ is an unbiased estimator of the high-fidelity model $s_1$, i.e., $\mathbb{E}[\bar{s}_{\rm{MLMC}}] = \mathbb{E}[s_1]$.
Its MSE is given by the formula
\begin{equation}\label{mlmcvar}
    \begin{aligned}
    \text{MSE}(\bar{s}_{\rm{MLMC}}) &= 
    \mathbb{E}[(\bar{s}_{\rm{MLMC}}-\mathbb{E}[s_1])^2]\\
    &= \frac{\sigma_L^2}{n_L} + \sum_{j=1}^{L-1} \frac{\levelvariance{j}{j+1}}{n_{j}}
\end{aligned}
\end{equation}
where we are using the auxiliary variable 
\begin{align*}
    \levelvariance{j}{j+1} := \mathbb{V}\text{ar}(s_{j}-s_{j+1}) = \Sigma_{j, j} - 2\Sigma_{j, j+1} + \Sigma_{j+1, j+1}
\end{align*}
to abbreviate the variance of each level.
The underlying premise for the MLMC estimator to be effective is the observation that if models $s_{j}$ and $s_{j+1}$ yield similar OoIs, then the level variance $\levelvariance{j}{j+1}$ is small; consequently only a small sample size $n_{j}$ is required to balance out the contribution of level $j$ to the MSE of the MLMC estimator in \eqref{mlmcvar}.
For level $L$, the surrogate $s_L$ is sampled alone without an additional model to decrease that level's variance, but since $s_L$ is the cheapest model it can be expected that the associated cost $c_L n_L$ remains reasonable, even if $n_L$ is large.

To balance and minimize the contributions of each level to the MLMC estimator's MSE,
it suggested in \cite{giles2015multilevel} to choose 
\begin{equation}\label{eq:mlmc:samplesizes}
    \begin{aligned}
    &n_L = \tau \sqrt{\frac{\sigma_L^2}{c_L}}, \quad
    n_{j} = \tau \sqrt{\frac{\levelvariance{j}{j+1}}{c_{j} + c_{j+1}}}
    \end{aligned}
\end{equation}
for $j = 1, \dots, L-1$.
The scaling factor $\tau>0$ can be chosen to adhere to budget constraints:
For $\bar{s}_{\rm{MLMC}}$ to have a cost of $c$, i.e., 
\begin{align*}
    n_L c_L + \sum_{j=1}^{L-1} n_{j}(c_{j}+c_{j+1}) = c,
\end{align*}
is equivalent to choosing 
\begin{align}\label{eq:mlmc:ratio}
    \tau=c \left(\sqrt{\sigma_L^2 c_L} + \sum_{j=1}^{L-1} \sqrt{\levelvariance{j}{j+1}(c_{j} + c_{j+1})} \right)^{-1}.
\end{align}
To obtain integers, sample sizes $n_{j} \le 1$ are rounded up to 1, and down otherwise.
Note that the rounding may cause the computational budget to be violated\footnote{To our knowledge there does not yet exist a low-budget sample size optimization for MLMC.
}; the budget constraint therefore needs to be checked before accepting a combination of surrogate models (see Step 2.4 in Algorithm \ref{alg:mlmc}).

\begin{algorithm}[t]
\textbf{Input:  }{
high-fidelity model $s_1$, 
surrogate models $s_2, \dots, s_L$, 
model evaluation costs $c_1 \ge \dots \ge c_L$, 
level variances $\levelvariance{1}{2}, \dots, \levelvariance{L-1}{L}$,
model variance $\sigma_L^2$ of lowest fidelity model,
computational budget $c$
}\\
\textbf{Output: }{sample sizes $1 \le n_1 \le \dots \le n_L$, estimator MSE $\texttt{MSE} = \mathbb{E}[(\bar{s}_{\rm{MFMC}}-\mathbb{E}[s_1])^2]$}
\begin{enumerate}
    \item Compute the scaling ratio $\tau$ using \eqref{eq:mlmc:ratio}.
    \item Compute $n_1, \dots, n_L \in \mathbb{R}$ with \eqref{eq:mlmc:samplesizes}.
    \item For $k=1, \dots, L$, if $n_k \le 1$ round it up to 1, otherwise round down. 
    \item Compute the MSE of the MLMC estimator with \eqref{mlmcvar} and save in variable \texttt{MSE}
    \item Return $(n_1, \dots, n_L)$, $\texttt{MSE}$
\end{enumerate}
\caption{Multilevel Monte Carlo}
\label{alg:mlmc}
\end{algorithm}

The computation of the MLMC sample sizes and corresponding MSE is summarized in Algorithm \ref{alg:mlmc}.
Similar to MFMC, the smallest MSE of the MLMC method may be realized by a subset of the available surrogate models.
Algorithm \ref{alg:mlmc} should hence be placed within an outer loop over all combinations of available surrogate models to determine the optimal model selection for which the returned value \texttt{MSE} is minimal.
If the identified MSE is smaller than $\sigma_1^2 / \lfloor c/c_1 \rfloor$ --- the MSE of MC sampling --- then the MLMC estimator can be computed using the chosen models and identified sample sizes using \eqref{eq:MLMC}.

\subsection{Multilevel Best Linear Unbiased Estimator}\label{sec:mlblue}

The MLBLUE method was introduced in \cite{Schaden2020} and has since been extended with theoretical results in \cite{schaden2021asymptotic} and with algorithms for sample size optimization and multiple OoIs in \cite{Croci2023}.
In the following, we focus on the main concept and algorithmic steps, and refer to the cited literature for details.
A full description of the MLBLUE method is provided in Algorithm \ref{alg:mlblue}.

To explain the structure of the MLBLUE estimator, we first define the vector of expectations of all available models
\begin{align*}
    \hat{\mathbf{s}} = \left(\mathbb{E}[s_1], \dots, \mathbb{E}[s_L]\right)^{\top} \in \mathbb{R}^{L}.
\end{align*}
Estimating the high-fidelity expectation $\mathbb{E}[s_1]$ is then equivalent to estimating $\mathbf{e}_1^{\top} \hat{\mathbf{s}}$ with $\mathbf{e}_1^{\top} = (1, 0, \dots, 0) \in \mathbb{R}^L$.
Let $\mathcal{S}_1, \dots, \mathcal{S}_{\hat{L}}$ be an enumeration of all $\hat{L} := 2^L$ non-empty subsets of the model indices $\{1, \dots, L\}$.

Focusing on any one index group $\mathcal{S}_i = \{1 \le j_1 \le \dots \le j_{L'} \le L\}$ with $L' := |\mathcal{S}_i|$, we define the random vector $\mathbf{s}_{i} := (s_{j_1}, \dots, s_{j_{\tilde{L}}})^{\top}$ by stacking the models $s_{j_k}$ with indices $j_k \in \mathcal{S}$ together.
The evaluation $\mathbf{s}_{i}(\theta)$ at a sample $\theta \sim \nu$ can then be interpreted as a noisy observation of $\hat{\mathbf{s}}$:
\begin{equation}\label{eq:MLBLUE:singleGroup}
    \begin{aligned}
    \mathbf{s}_{i}(\theta)
    &= (s_{j_1}(\theta), \dots, s_{j_{\tilde{L}}}(\theta))^{\top} \\
    &= \left(\mathbb{E}[s_{j_1}], \dots, \mathbb{E}[s_{j_{\tilde{L}}}] \right)^{\top} \\
    &\quad + \left(s_{j_1}(\theta)-\mathbb{E}[s_{j_1}], \dots, s_{j_{\tilde{L}}}-\mathbb{E}[s_{j_{\tilde{L}}}] \right)^{\top} \\
    &=: \mathbf{R_{i}} \hat{\mathbf{s}} + \varepsilon_{i}(\theta)
\end{aligned}
\end{equation}
where the matrix $\mathbf{R_{i}} \in \{0, 1\}^{L' \times L}$ removes all entries in $\hat{\mathbf{s}}$ that are not in the index set $\mathcal{S}_i$; it is defined through $(\mathbf{R_{i}})_{k,\ell} = \delta_{\ell, j_k}$ using the Kronecker delta.
By construction, $\varepsilon_{\mathcal{S}} \in \mathbb{R}^{L'}$ is a random variable with mean zero, and covariance matrix $\mathbf{C}_{i} \in \mathbb{R}^{L' \times L'}$ defined by $(\mathbf{C}_{i})_{k, \ell} := \Gamma_{j_k, j_{\ell}}$.
Since $\mathbf{C}_{i}$ is a submatrix of the symmetric positive definite covariance matrix $\Gamma$ it is itself symmetric positive definite and thus invertible.

For a given vector $\mathbf{n} \in \mathbb{N}^{\hat{L}}_{\ge 0}$ of sample sizes for each model index group, we evaluate the random vector $\mathbf{s}_i$ at $\mathbf{n}_i$ i.i.d. samples $\left\{\theta_{k}^{(i)}\right\}_{k=1}^{\mathbf{n}_i}$.
An estimate of $\hat{\mathbf{s}}$ is then obtained by solving the regression problem
\begin{align*}
    \min_{\mathbf{s} \in \mathbb{R}^L} \sum_{i=1}^{\hat{L}} \sum_{k=1}^{n_i} \left( \mathbf{R}_{i} \mathbf{s} - \mathbf{s}_{i}(\theta_{k}^{(i)})\right)^{\hspace{-0.3em}\top} \hspace{-0.3em} \mathbf{C}_{i}^{-1} \hspace{-0.3em} \left( \mathbf{R}_{i} \mathbf{s} - \mathbf{s}_{i}(\theta_{k}^{(i)}) \right)
\end{align*}
The minimum norm solution to this problem is obtained at
\begin{align}\label{eq:MLBLUE:minnorm}
    \hat{\mathbf{s}}^* = \Psi(\mathbf{n})^{\dagger} \mathbf{y}(\mathbf{n})
\end{align}
with the vector 
\begin{align}\label{eq:MLBLUE:y}
    \mathbf{y}(\mathbf{n}) &:= \sum_{i=1}^{\hat{L}} \mathbf{R}_{i} \mathbf{C}_{i}^{-1} \sum_{k=1}^{n_i} \mathbf{s}_{i}(\theta_k^{(i)}) \in \mathbb{R}^{L}
\end{align}
and using the Moore-Penrose pseudoinverse of the likelihood matrix
\begin{align}\label{eq:MLBLUE:likelihood}
    \Psi(\mathbf{n}) &:= \mathbf{R}(\mathbf{n}) \mathbf{C}_{\varepsilon}(\mathbf{n})^{-1} \mathbf{R}(\mathbf{n})^{\top} \in \mathbb{R}^{L \times L}.
\end{align}
In \eqref{eq:MLBLUE:likelihood}, the matrices $\mathbf{R}(\mathbf{n}) \in \{0, 1\}^{L \times m}$ and $\mathbf{C}_{\varepsilon}(\mathbf{n}) \in \mathbb{R}^{m \times m}$, with $m = m(\mathbf{n}) := \sum_{i=1}^{\hat{L}}\mathbf{n}_i|\mathcal{S}_{i}|$, contain $\mathbf{n}_i$ copies of each matrix $\mathbf{R}_i$ and $\mathbf{C}_i$:
\begin{equation}\label{eq:MLBLUE:def}
\begin{aligned}
     \mathbf{R}(\mathbf{n}) &:= (\mathbf{R}_{1}, \dots, \mathbf{R}_{1}, R_{2}, \dots, \mathbf{R}_{{\hat{L}}}), \\
     \mathbf{C}_{\varepsilon}(\mathbf{n}) &:= \text{diag}(\mathbf{C}_{1}, \dots, \mathbf{C}_{1}, \mathbf{C}_{2}, \dots, \mathbf{C}_{\hat{L}}). 
\end{aligned} 
\end{equation}
After solving \eqref{eq:MLBLUE:minnorm}, we obtain the MLBLUE estimator
\begin{align}\label{eq:MLBLUE:estimator}
    \bar{s}_{\rm{MLBLUE}} := \mathbf{e}_1^{\top} \hat{\mathbf{s}}^* \approx \mathbb{E}[s_1].
\end{align}

\noindent It was shown in \cite{Schaden2020, gross2004general}, that $\bar{s}_{\rm{MLBLUE}}$ is an unbiased estimator, i.e. $\mathbb{E}[\bar{s}_{\rm{MLBLUE}}] = \mathbb{E}[s_1]$, if the high-fidelity model $s_1$ is sampled at least once.
This condition can be written equivalently as $\mathbf{n}^{\top}\mathbf{h} \ge 1$ where $\mathbf{h} \in \{0, 1\}^{\hat{L}}$ is defined by $\mathbf{h}_i = \delta_{1 \in \mathcal{S}_i}$.
In this case, the MLBLUE estimator's MSE is given by 
\begin{equation}\label{eq:MLBLUE:var}
    \begin{aligned}
    \text{MSE}(\bar{s}_{\rm{MLBLUE}})
    &= \mathbb{E}[(\bar{s}_{\rm{MLBLUE}}-\mathbb{E}[s_1])^2] \\
    &= \mathbf{e}_1^{\top} \Psi(\mathbf{n})^{\dagger} \mathbf{e}_1.
\end{aligned}
\end{equation}

The sample size $\mathbf{n}_i$ for each model vector $\mathbf{s}_i$ should be chosen to minimize the MSE of the MLBLUE estimator $\bar{s}_{\rm{MLBLUE}}$.
It was shown in \cite{Croci2023} that the optimal $\mathbf{n}$ for a given budget $c$ solves the semi-definite programming problem
\begin{equation}\label{eq:SDP}
    \begin{aligned}
    &\min_{\mathbf{n}\ge 0, t\in \mathbb{R}} t \\
    & \quad \text{s.t. }
    \left(
    \begin{array}{cc}
        \Psi(\mathbf{n}) & \mathbf{e}_1 \\
        \mathbf{e}_1^{\top} & t
    \end{array}
    \right) \text{ is s.p.d.,}\\
    & \quad \text{and }
    \mathbf{n}^{\top}\mathbf{c} \le c, ~\mathbf{n}^{\top}\mathbf{h} \ge 1,
\end{aligned}
\end{equation}
where the acronym s.p.d. stands for symmetric positive definite, and the vector $\mathbf{c} \in \mathbb{R}^{\hat{L}}$ contains the costs for evaluating the model group $i$, i.e., $\mathbf{c}_i = \sum_{j \in \mathcal{S}_i} c_j \ge 0$.
The obtained sample size vector $\mathbf{n}$ can then be used to compute the optimal MLBLUE estimator $\bar{\mathbf{s}}_{\rm{MLBLUE}}$.
The procedure is summarized in Algorithm \ref{alg:mlblue}.

\begin{algorithm}[t]
\textbf{Input:  }{
high-fidelity model $s_1$, 
surrogate models $s_2, \dots, s_L$, 
model evaluation costs $c_1 \ge \dots \ge c_L$, 
model covariance matrix $\Sigma \in \mathbb{R}^{L\times L}$, 
computational budget $c$}\\
\textbf{Output: }{
MLBLUE estimator $\bar{s}_{\rm{MLBLUE}} \approx \mathbb{E}[s_1]$, 
estimator variance $\mathtt{MSE} = \mathbb{V}\text{ar}(\bar{s}_{\rm{MLBLUE})}$}
\begin{enumerate}
    \item Set $\hat{L} := 2^L$.
    Choose an enumeration $\mathcal{S}_1, \dots, \mathcal{S}_{\hat{L}}$ of all $\hat{L}$ non-empty subsets of the model indices $\{1, \dots, L\}$.
    \item For $i=1, \dots, \hat{L}$, compute the matrices $\mathbf{C}_i$ and $\mathbf{R}_i$ (defined after \eqref{eq:MLBLUE:singleGroup}), and the cost entry $\mathbf{c}_i = \sum_{k=1}^{|\mathcal{S}_i|} c_{j_k}$ for the model index set $\mathcal{S}_i = \{j_1 \le \dots \le j_{|\mathcal{S}_i|}\}$
    \item Using the definitions in \eqref{eq:MLBLUE:likelihood} and \eqref{eq:MLBLUE:def}, solve the semi-definite programming problem \eqref{eq:SDP} to obtain the sample sizes $\mathbf{n}$. We suggest using an off-the-shelf solver.
    \item To compute the MLBLUE MSE, first compute $\mathbf{R}(\mathbf{n})$ and $\mathbf{C}_{\varepsilon}(\mathbf{n})$ with \eqref{eq:MLBLUE:def}, and use them to compute $\Psi(\mathbf{n})$ via \eqref{eq:MLBLUE:likelihood}.
    Then evaluate $\mathtt{MSE} = \mathbf{e}_1^{\top} \Psi(\mathbf{n})^{\dagger} \mathbf{e}_1$ using \eqref{eq:MLBLUE:var}.
    \item To compute the MLBLUE estimator, for each $1\le i\le \hat{L}$ with $\mathbf{n}_i \neq 0$, draw samples $\theta_k^{(i)} \sim \nu$, $1 \le k \le \mathbf{n}_i$, and evaluate $s_j(\theta_k^{(i)})$ for each model index $j \in \mathcal{S}_i$.
    Use these output evaluations in \eqref{eq:MLBLUE:y} to compute $\mathbf{y}(\mathbf{n})$.
    Then compute the minimum norm solution $\hat{\mathbf{s}}^*$ of $\Psi(\mathbf{n}) \mathbf{s} = \mathbf{y}(\mathbf{n})$ using \eqref{eq:MLBLUE:minnorm}.
    Finally, compute the MLBLUE estimator $\bar{\mathbf{s}}_{\rm{MLBLUE}}$ using \eqref{eq:MLBLUE:estimator}.
    \item return $\bar{\mathbf{s}}_{\rm{MLBLUE}}$, \texttt{MSE}
\end{enumerate}
\caption{Multilevel Best Linear Unbiased Estimator}
\label{alg:mlblue}
\end{algorithm}

Both the MFMC and the MLMC estimator can be written in the MLBLUE index group structure, albeit with pre-determined weights. 
Recall that MLMC optimizes the sample size and MFMC optimizes both the sample size and the weights.
The MLBLUE method goes further to optimize sample sizes jointly with the estimator's structure in the form of selected model groups and their weights. This means that by construction, the MSE of the MLBLUE estimator is at least as small as that of the MFMC and the MLMC method, meaning that the MLBLUE estimator is guaranteed to be at least as good as MFMC's and MLMC's, and in many cases will be better.
However, since the number $\hat{L} = 2^L$ of model index groups grows exponentially fast in the number $L-1$ of available surrogate models and because the covariance matrix $\Gamma$ may be arbitrarily badly conditioned and biased with approximation errors, solving \eqref{eq:SDP} and \eqref{eq:MLBLUE:minnorm} can become numerically challenging.
In particular, the MLBLUE optimization tends to select dissimilar model index groups, sample sizes, and weights even for similar computational budgets.
In contrast, the MFMC method has been shown to be robust towards approximation errors in $\Gamma$ (see \cite{Peherstorfer2016b}).
Since MLBLUE has only recently been introduced, similar advancements have not yet been made, though they can likely be expected from future work.

\section{Application to Greenland mass loss projections}\label{sec:results}

We conclude this paper with a demonstration of the introduced multifidelity UQ methods MFMC, MLMC, and MLBLUE.
Specifically, we estimate the expected ice mass loss of the Greenland ice sheet high-fidelity model under uncertainties in the basal friction field and the geothermal heat flux for the 2015--2050 time period.
Our surrogate modeling setup is described in Section \ref{sec:setup}.
For our demonstration, we distinguish between two UQ use cases:
First, in Section \ref{sec:results:accuracy} we prescribe a target accuracy in the estimated mean of $\pm \SI{1}{mm}$ SLR equivalent (or \SI{\pm361.8}{Gt}) ice mass loss at a \SI{95}{\%} confidence level;
second, in Section \ref{sec:results:budget}, we compare the accuracy of the 2015--2050 estimates obtained for a fixed computational budget of five high-fidelity model evaluations.
We conclude with a discussion of the results in Section \ref{sec:discussion}.

\subsection{Surrogate models}\label{sec:setup}

In our multifidelity modeling setting, we employ $L=13$ models of the ice mass change in Greenland for the years 2015--2050.
Our high-fidelity model $s_1$ is governed by the equations for ice temperature, ice thickness, and the HO stress balance model for the ice velocity as described in Section~\ref{sec:high-fidelity}.
It is discretized on our finest available mesh, denoted ``fine.''
Surrogate models $s_3$ and $s_5$ (named ``HO, medium'' and ``HO, coarse'', respectively) are obtained by replacing the high-fidelity mesh with the two coarser meshes denoted ``medium'' and ``coarse,'' with initial conditions obtained by interpolating the $t=2015$ high-fidelity initial condition followed by a relaxation run.
The mesh resolutions are provided in Table~\ref{tab:discretization}.
Surrogate models $s_7$ (``SSA, fine"), $s_9$ (``SSA, medium"), and $s_{11}$ (``SSA, coarse"), are obtained by replacing the HO stress balance solver with the SSA, which approximates the three-dimensional velocity fields $\velocity_x$ and $\velocity_y$ by their two-dimensional depth-averages (see Section~\ref{sec:surrogates}).
All models are run using ISSM with adaptive time stepping adhering to a CFL condition to guarantee numerical stability.
The average computational costs are listed in Table~\ref{tab:surrogates}.
We did not encounter any model blow-ups, and all runs concluded without errors and warnings.

\begin{table}[]
    \centering
    \begin{tabular}{l|r|lll}
    \toprule
        model & cost & \multicolumn{3}{c}{model setup} \\
        no. & [CPU-h] & type & mesh & velocity \\
    \midrule
    1 & 64.888 & ISSM & fine & HO \\
    2 & 37.079 & extr. & fine & HO \\
    3 & 22.889 & ISSM & medium & HO \\
    4 & 13.079 & extr. & medium & HO \\
    5 & 4.689 & ISSM & coarse & HO \\
    6 & 2.679 & extr. & coarse & HO \\
    7 & 0.455 & ISSM & fine & SSA \\
    8 & 0.260 & extr. & fine & SSA \\
    9 & 0.105 & ISSM & medium & SSA \\
    10 & 0.060 & extr. & medium & SSA \\
    11 & 0.059 & ISSM & coarse & SSA \\
    12 & 0.034 & extr. & coarse & SSA \\
    13 & $<0.001$ & interp. & coarse & SSA \\
    \botrule
    \end{tabular}
    \caption{Surrogate model overview}
    \label{tab:surrogates}
\end{table}

\begin{figure*}
    \centering
\includegraphics[width=\textwidth]{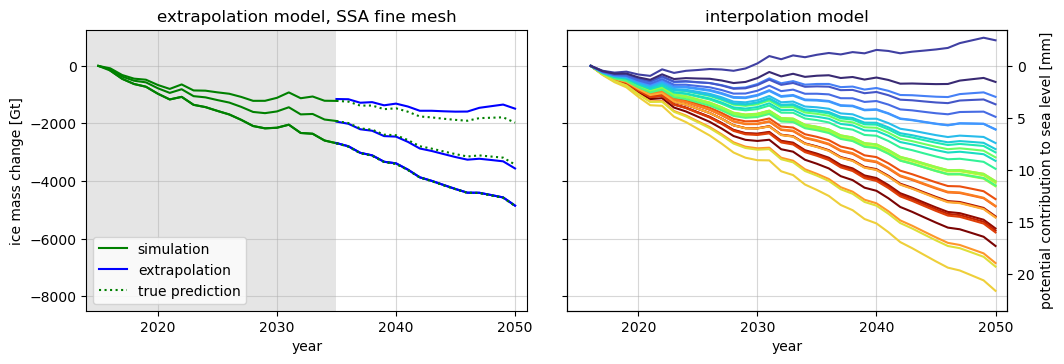}
    \caption{Left: schematic explanation of extrapolation surrogate modeling setup. Right: Interpolation-based predictions for 32 samples of the geothermal heat flux and basal friction fields}
    \label{fig:extrapolation}
\end{figure*}

For each of the ISSM models $s_1$, $s_3$, $s_5$, $s_7$, $s_9$, and $s_{11}$, we define data-fit surrogate models that we denote ``extrapolation'' data-fit model. 
These data fit surrogates are denoted
$s_2$, $s_4$, $s_6$, $s_8$, $s_{10}$, and $s_{12}$,
and are based on the observation that the ice mass loss is dominated by two trends, 
the overall, almost linear decline whose slope varies by parameter samples, and yearly fluctuations.
The latter is primarily caused by the surface mass balance $\dot{M}_{s}$, and only implicitly depends on the parameter samples through the coupling of $\dot{M}_{\rm{s}}$ to the ice altitude, and the coupling of the basal melt $\dot{M}_{\rm{b}}$ to the geothermal heat flux.
Both trends can clearly be observed in Figures~\ref{fig:predictions} (left) and \ref{fig:predictions:2x3}.
To exploit this effect, we first compute nine reference samples of the ``SSA, coarse'' models, fit a linear function to each, and take the mean over the misfit to approximate the parameter-independent yearly adjustments.
We then define ``extrapolation'' surrogate models by first running our HO or SSA models for 20 years using a sampled parameter, and then fitting a linear curve and the learned yearly adjustments to the thus obtained 2015-2035 predictions;
to obtain predictions for $t\in[2035, 2050]$, we then evaluate the obtained data-fit model.
The procedure is illustrated in Figure~\ref{fig:extrapolation}.
As the cost is dominated by the first simulation step in ISSM, each extrapolation model's predictions for 2015--2050 are $35/20 = 1.75$ times cheaper than its corresponding ISSM model.

Finally, for our cheapest surrogate model $s_{13}$ we interpolate linearly between the nine reference ice mass change predictions of ``SSA, coarse.''
This interpolation is extremely cheap  but is --- unsurprisingly --- not particularly accurate for individual predictions, c.f.\ Figure~\ref{fig:extrapolation}.
Still, as will be seen in the following results, the interpolation model achieves a high correlation 
with the high-fidelity model for the 2050 ice mass loss, making it beneficial to employ in the multifidelity estimators.
Thus, this interpolation model demonstrates the benefit of cheap surrogate models even at the cost of reduced accuracy, when employed within a formal multifidelity estimation framework.

The multifidelity UQ methods require an estimate of the model covariance matrix $\Sigma \in \mathbb{R}^{13\times13}$.
We use Latin Hypercube sampling to choose 32 parameter samples for computing all entries in $\Sigma$ that depend on HO models.
The ISSM model predictions for the parameters are shown in Figure \ref{fig:predictions:2x3}, and for the high-fidelity model additionally in Figure \ref{fig:predictions}.
We do not remove outliers from the sampled data (as seen in Figure \ref{fig:predictions:2x3} for the ``SSA, medium" model) to avoid estimating an overly confident covariance matrix.
For all entries in $\Sigma$ that do not depend on the HO models but on the cheaper SSA models, we include an additional 96 parameters in the sampling to reduce potential bias in $\Sigma$.
Finally, for $\Sigma_{13,13}$, the variance of the interpolation model $s_{13}$, we use $50,000$ samples.
We record the model costs $c_1, \dots, c_{13}$ as the average CPU time for computing the predictions at these samples.
Note that we do not count these computations towards our budget for estimating the high-fidelity expectation because we expect that for state-of-the-art ice sheet models, $\Sigma$ can be approximated from verification and validation runs performed during model development, or using convergence rates.
If not, then the pilot samples computed for its estimation could also be reused in the multifidelity estimators.

\subsection{Target accuracy}\label{sec:results:accuracy}

\begin{figure*}
    \flushright
\includegraphics[width=0.9\textwidth]{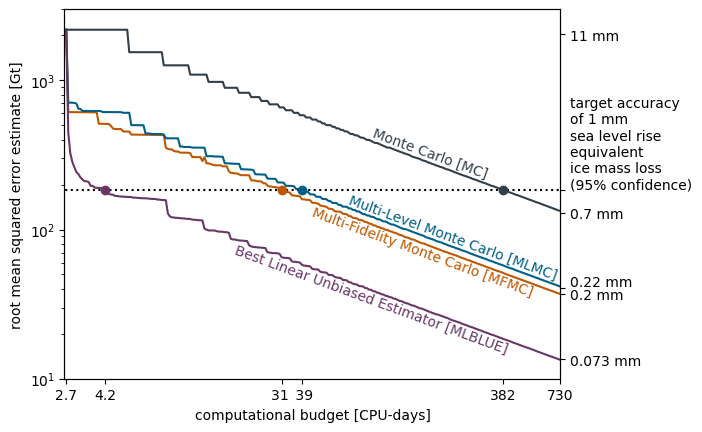}
    \caption{Comparison of root MSE of the multifidelity UQ methods MFMC, MLMC, and MLBLUE, and Monte Carlo sampling for any given computational budget 
    }
    \label{fig:cost-vs-accuracy}
\end{figure*}

We start by comparing the MSE that can be achieved by MFMC, MLMC, and MLBLUE for different computational budgets when approximating the expected 2050 ice mass loss, see Figure \ref{fig:cost-vs-accuracy}.
Based on the heterogeneity of available models, the MSE for MFMC is slightly smaller than that of MLMC for the same budget, though both are larger than the MSE achieved by MLBLUE.
Asymptotically, the MSE for the MFMC estimator is 13.0 times smaller than that of Monte Carlo for the same budget.
Similarly, MLMC reduces the MSE by a factor of 10.2, and MLBLUE by a factor of 98.1.
Interpreted in terms of accuracy, this means that the multifidelity estimators achieve a computational speedup between one and two orders of magnitude compared to MC sampling.

We compare the performance of the MFMC, MLMC, and MLBLUE methods when models and sample sizes are chosen to obtain a prescribed target accuracy of \SI{\pm 361.8}{Gt}, corresponding to \SI{\pm1}{mm} SLR equivalent ice mass loss, at a \SI{95}{\%} confidence level.
The computational budget required by the multifidelity methods to achieve this target accuracy is 30.14 CPU-days for MFMC, 37.52 CPU-days for MLMC, and 4.19 CPU-days for MLBLUE.
These values are marked with dots in Figure \ref{fig:cost-vs-accuracy}.
Compared to the 381.22 CPU-days required by Monte Carlo sampling, these correspond to computational speed-ups of factors 12.6 for MFMC, 10.2 for MLMC, and 91.0 for MLBLUE.
Note that the speedup for MLBLUE is smaller than its expected asymptotic value ($98.1\times$) because the estimator is falling into the low-budget regime (see Figure~\ref{fig:cost-vs-accuracy}).

\begin{figure*}
    \flushleft
\includegraphics[width=\textwidth]{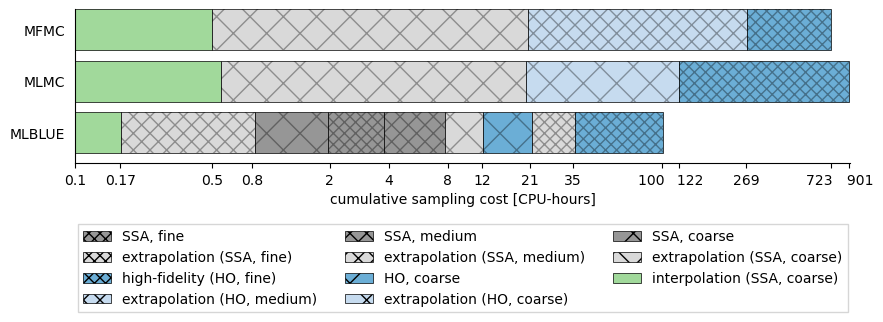}
    \caption{Distribution of computational budget for MFMC, MLMC, and MLBLUE for reaching a target accuracy of \SI{\pm 361.8}{Gt} at \SI{95}{\%} confidence
    }
    \label{fig:samplesizes}
\end{figure*}

The optimal distribution of surrogate models and sampling costs for achieving the target accuracy are shown in Figure \ref{fig:samplesizes}.
All three methods rely on information from all four surrogate model types (mesh coarsening, physical approximation, extrapolation, and interpolation).
The high-fidelity model is sampled 7 times by MFMC, 12 by MLMC, and only once by MLBLUE, illustrating how all estimators are able to shift the computational burden onto the surrogate models.
MFMC and MLMC rely heavily on the computationally cheap but least accurate interpolation model with more than 180,000 samples each.
In contrast, MLBLUE uses the interpolation model far less (61,573 samples), and opts for a larger variety in models.
In particular, MLBLUE uses all six SSA surrogate models (both full 2015--2050 ISSM predictions and their extrapolations, each for the three available meshes) and the full 2015--2050 HO-coarse, while MFMC and MLMC only use extrapolation and interpolation type surrogates.

\begin{table*}
    \centering
    \begin{tabular}{ll|cc|cc|r}
    \toprule
        & & \multicolumn{2}{c|}{ice mass loss [Gt]} & \multicolumn{2}{c|}{SLR contribution [mm]} & cost \\
        setting & method & estimate & uncertainty & estimate & uncertainty & [CPU-days] \\
        \midrule
        & MC & -5511 & \hphantom{1} $\pm  360$ & 15.2 & $\pm 1.00$ & 381.2 \\
        target & MFMC & -5868 & \hphantom{1}$\pm 356$ & 16.2 & $\pm0.98$ & 30.1 \\
        accuracy & MLMC & -5757 & \hphantom{1}$\pm352$ & 15.9 & $\pm0.97$ & 37.5 \\
        & MLBLUE & -5725 & \hphantom{1}$\pm346$ & 15.8 & $\pm 0.96$ & 4.2 \\
        \midrule
        reference & MC & -5618 & \hphantom{1}$\pm 195$ & 15.5 & $\pm 0.54$ & 1297.8 \\
        \midrule 
        & MLBLUE & -5572 & \hphantom{1}$\pm 193$ & 15.4 & $\pm 0.53$ & 13.5 \\
        target & MLMC & -5639 & \hphantom{1}$\pm 608$ & 15.6 & $\pm 1.68$ & 12.6 \\
        budget & MFMC & -5581 & \hphantom{1}$\pm 612$ & 15.4 & $\pm1.69$ & 11.8 \\
        & MC & -5586 & $\pm 1913$ & 15.4 & $\pm 5.29$ & 13.5 \\
        \botrule
    \end{tabular}
    \caption{Expected Greenland ice mass loss by 2050: estimates obtained by MC and multifidelity UQ methods and associated uncertainty at a \SI{95}{\%} confidence level }
    \label{tab:predictions}
\end{table*}

\begin{figure*}
    \centering
    \includegraphics[width=\textwidth]{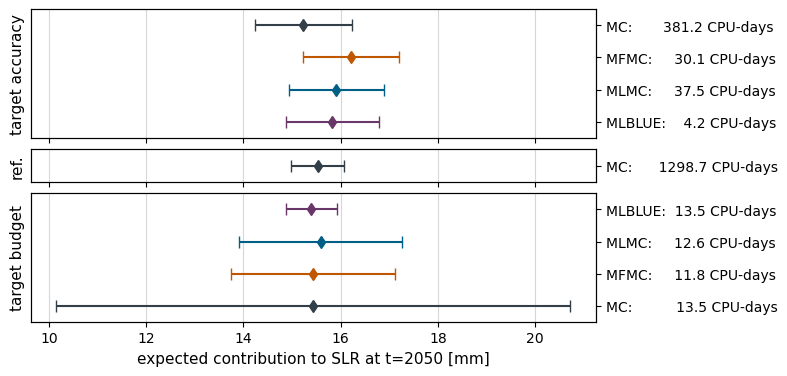}
    \caption{Confidence intervals for expected Greenland ice mass change in 2050 relative to 2015 and control run}\label{fig:confidenceintervals:acc}
\end{figure*}

We next approximate the expected high-fidelity ice mass loss for the year 2050 $\mathbb{E}[s_1(2050)]$ with each multifidelity UQ method, using the optimal sample sizes from Figure~\ref{fig:samplesizes} to guarantee an accuracy of at most \SI{\pm 1}{mm} SLR equivalent ice mass loss at a \SI{95}{\%} confidence level.
The predictions are provided in Table \ref{tab:predictions}, with confidence levels illustrated in Figure~\ref{fig:confidenceintervals:acc}.
For comparison, we also provide Monte Carlo predictions at the same accuracy, and at $\SI{\pm 0.54}{mm}$ SLR equivalent ice mass loss (at \SI{95}{\%} confidence).
All five confidence intervals overlap on the interval $[15.23, 16.07]$.

\subsection{Target budget}\label{sec:results:budget}

\begin{figure*}
    \centering
    \includegraphics[width=\textwidth]{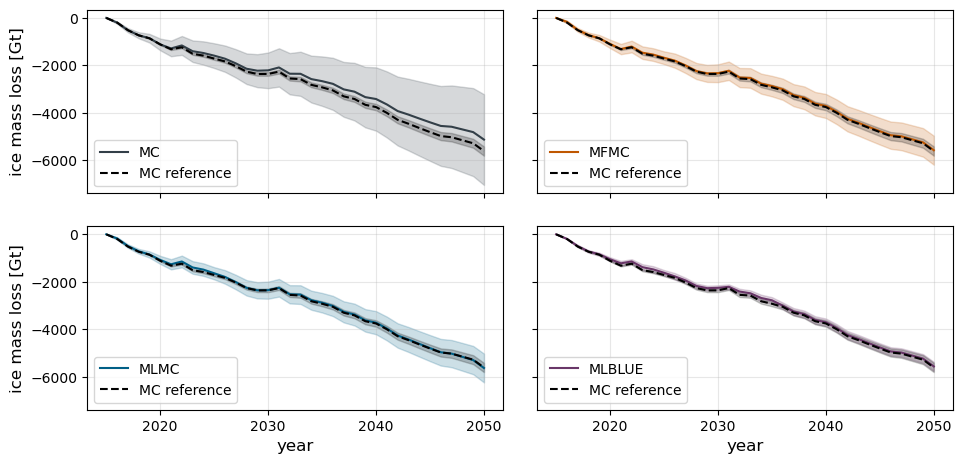}
    \caption{
    Estimates of the 2015-2050 ice mass change and 95\% confidence interval (shaded) for a computational budget of 324.44 CPU-h}
    \label{fig:accuracytest:budget}
\end{figure*}

We next demonstrate the power of multifidelity UQ when working with a restrictive computational budget.
For a budget $c = 13.5$ CPU-days, corresponding to five solves of the high-fidelity model, we compute one instance of each multifidelity estimator and its 95\% confidence interval for predicting the expected 2015--2050 ice mass change.
The predictions are shown in Figure \ref{fig:accuracytest:budget}.
The uncertainty in all estimators increases in time, with Monte Carlo sampling performing the worst --- as expected  for the small number of high-fidelity samples.
Its final 95\% confidence interval at $t=2050$ is $[-5132.9 \pm 1913.1]$ Gt, or $[14.19 \pm 5.29]$ mm SLR equivalent ice mass loss, as illustrated in Figure~\ref{fig:confidenceintervals:acc}.
In contrast, the final confidence intervals for MFMC, MLMC, and MLBLUE are $[-5581.1\pm 611.6]$ Gt, $[-5639.4 \pm 607.7]$ Gt, and $[-5572.2 \pm 193.1]$ Gt, corresponding to an accuracy of $\pm 1.69$, $\pm 1.68$, and $\pm 0.53$ mm SLR equivalent ice mass loss at $95\%$ confidence.
For Monte Carlo sampling to achieve these accuracies at this level of confidence, we would require a computational budget of 132.5, 135.2 and 1327.5 CPU-days (49, 50, and 491 high-fidelity model solves).

We note that for Figure \ref{fig:accuracytest:budget} the sample sizes and weights of the multifidelity UQ estimators were chosen to minimize the error incurring throughout the 2015--2050 prediction regime.
This stands in contrast to the values reported in Figure \ref{fig:cost-vs-accuracy}, where the estimators were optimized for the single prediction at $t=2050$.
In particular, in Figure \ref{fig:accuracytest:budget}, MLMC performed better than MFMC with up to 5\% smaller confidence intervals.
However, after rounding the sample sizes, neither method fully exploited the computational budget but only used \SI{87.348}{\%} (MFMC) and \SI{92.882}{\%} (MLMC), while MLBLUE used \SI{99.998}{\%}.
Thus, the performance of both methods can still be improved by redistributing the remaining computational budget, e.g., via \cite{Gruber2023}.

\subsection{Conclusion}\label{sec:discussion}

In this paper we have explained three multifidelity UQ techniques --- MFMC, MLMC, and MLBLUE --- and applied them to compute the expected 2050 ice mass loss of the Greenland ice sheet.
Despite strong variability in individual high-fidelity predictions caused by uncertain basal friction and geothermal heat flux fields, the multifidelity UQ estimators achieve an approximation accuracy of \SI{\pm 1}{mm} SLR equivalent ice mass loss at a \SI{95}{\%} confidence level with up to $91\times$ speed-up compared to Monte Carlo sampling.
In the low-budget regime, the MLBLUE estimate for a computational budget of five high-fidelity solves has a variance that would require 491 Monte Carlo samples.
Overall, the three multifidelity UQ estimators have proven to be well-suited for computing the output expectations of highly expensive ice sheet simulations.

In our numerical experiments, MFMC and MLMC have performed similarly well to each other but have been outperformed by MLBLUE.
This result was expected as MLBLUE guarantees by construction an estimator variance that is smaller or equal to that of both the MFMC and the MLMC estimator.
We still recommend both MFMC and MLMC for UQ in ice sheet simulations, primarily because both methods have a strong foundation in the literature with extensions beyond forward UQ (e.g., Markov chain Monte Carlo and sensitivity analysis).
In contrast, similar extensions for MLBLUE are still in active development.
Moreover, while MFMC and MLMC are generally robust against against approximation bias in the input covariance matrix, the MLBLUE sample size optimization can be challenging.
Our recommended procedure after computing the input covariance matrix is to first evaluate the potential benefit of each method (e.g., in the form of Figure \ref{fig:cost-vs-accuracy}) before committing to one.
In either case, the results show clearly that the existing surrogate models in the ice sheet literature are sufficient to enable UQ even for highly expensive high-fidelity ice sheet models.

Naturally, the performance of the multifidelity UQ methods depends on the high-fidelity model and its surrogates.
We have chosen our modeling setup here to be as close to the ISMIP6 protocol \cite{nowicki2020experimental} as possible while still permitting comparisons with Monte Carlo sampling to allow an easy transfer of techniques to other models and codes.
In addition, we have employed surrogate models that are readily available and do not require expert-level surrogate modeling implementations.
We consequently might expect even better performance of the multifidelity UQ methods if specialized reduced-order models are employed among the surrogates.

\section*{Acknowledgements}
The authors would like to thank Matteo Croci for fruitful discussions and support using his code basis.
We acknowledge the World Climate Research Programme, which, through its Working Group on Coupled Modelling, coordinated and promoted CMIP6. We thank the climate modeling groups for producing and making available their model output, the Earth System Grid Federation (ESGF) for archiving the data and providing access, and the multiple funding agencies who support CMIP6 and ESGF.

\section*{Declarations}

\subsection{Funding}
This work was supported in parts by the Department of Energy grants DE-SC0021239 and DE-SC002317, and the Air Force Office of Scientific Research grant FA9550-21-1-0084. MM was funded by National Science Foundation grant \#2118285.

\subsection{Competing interests}
The authors have no competing interests to declare that are relevant to the content of this article.

\subsection{Research data availability}
The BedMachine v5 dataset on the Greenland bedrock topography and ice thickness is available at \href{https://nsidc.org/data/idbmg4/versions/5}{nsidc.org/data/idbmg4/versions/5}.
The GIMP ice and ocean mask (v2.0) is available at \\
\href{https://byrd.osu.edu/research/groups/glacier-dynamics/data/icemask}{byrd.osu.edu/research/groups/glacier-dynamics/data/icemask}.
Its combination with the coastline by Jeremie Mouginot is available at 
\url{issm.jpl.nasa.gov/documentation/tutorials/datasets/}
(SeaRISE Greenland dev1.2).
The MEaSUREs Multi-year Greenland Ice Sheet Velocity Mosaic (version 1) is available at \href{https://nsidc.org/data/nsidc-0670/versions/1}{nsidc.org/data/nsidc-0670/versions/1}.
The MEaSUREs Greenland Annual Ice Sheet Velocity Mosaics from SAR and Landsat (version 4, ID NSIDC-0725) for the years 2015-2021 are available at \href{https://nsidc.org/data/nsidc-0725/versions/4}{nsidc.org/data/nsidc-0725/versions/4}.
The CNRM-CM6-1 atmospheric projections are available on \href{https://theghub.org/}{theghub.org/}.
The Greenland heat flux A20180227-001 is available at \\
\href{https://ads.nipr.ac.jp/data/meta/A20180227-001/}{ads.nipr.ac.jp/data/meta/A20180227-001/}.
The Greenland heat flux ``Shapiro-Ritzwoller" is available at \href{http://ciei.colorado.edu/~nshapiro/MODEL/}{ciei.colorado.edu/$\sim$nshapiro/MODEL/}.

\subsection{Code availability}
The code and prediction data used to generate the results in this study is available at \\
\href{https://github.com/nicolearetz/Multifidelity-UQ-Greenland}{github.com/nicolearetz/Multifidelity-UQ-Greenland}.
The implementations of MFMC, MLMC, and MLBLUE are available at \href{https://github.com/croci/bluest}{github.com/croci/bluest}.
The Ice-sheet and Sea-level System Model is available at \href{https://github.com/ISSMteam/ISSM}{https://github.com/ISSMteam/ISSM}.

\begin{appendices}

\end{appendices}

\bibliographystyle{plain}
\bibliography{references}

\end{document}